\documentclass{aa}
\usepackage[utf8]{inputenc}
\usepackage{txfonts}
\usepackage{xspace}
\usepackage{microtype}
\usepackage[colorlinks=true,linkcolor=red,urlcolor=black,citecolor=blue]{hyperref}
\usepackage{tablefootnote}
\usepackage{csquotes}
\usepackage{multirow}

\title{An in-depth analysis of the variable cyclotron lines in \gx{}}
\author{
        Nicolas~Zalot\inst{\ref{inst:remeis}} \and
        Ekaterina~Sokolova-Lapa\inst{\ref{inst:remeis}}\and 
        Jakob~Stierhof \inst{\ref{inst:remeis}} \and
        Ralf~Ballhausen \inst{\ref{inst:umd},\ref{inst:gsfc}} \and 
        Aafia~Zainab \inst{\ref{inst:remeis}} \and
        Katja~Pottschmidt \inst{\ref{inst:gsfc},\ref{inst:CRESST}} \and
        Felix~F\"urst \inst{\ref{inst:esac}} \and
        Philipp~Thalhammer \inst{\ref{inst:remeis}} \and
        Nazma~Islam \inst{\ref{inst:gsfc},\ref{inst:CSST}} \and
        Camille~M.~Diez \inst{\ref{inst:esac}} \and
        Peter~Kretschmar \inst{\ref{inst:esac}} \and
        Katrin~Berger \inst{\ref{inst:remeis}} \and 
        Richard~Rothschild \inst{\ref{inst:ucsd}} \and
        Christian~Malacaria \inst{\ref{inst:issi}} \and
        Pragati~Pradhan \inst{\ref{inst:erau}} \and
        J\"orn~Wilms\inst{\ref{inst:remeis}}}
\authorrunning{Nicolas Zalot et al.}

\institute{
    Dr.\ Karl-Remeis-Observatory and Erlangen Centre for Astroparticle Physics,
    Friedrich-Alexander-Universit\"at Erlangen-N\"urnberg,
    Sternwartstr.~7, 96049 Bamberg, Germany \email{nicolas.zalot@fau.de} \label{inst:remeis}
\and
    University of Maryland, Department of Astronomy, College Park, MD 20742, USA \label{inst:umd}
\and 
    NASA Goddard Space Flight Center, Astrophysics Science Division, Greenbelt, MD 20771, USA \label{inst:gsfc}
\and
    CRESST and Center for Space Sciences and Technology, University of Maryland, Baltimore County, 1000 Hilltop Circle, Baltimore, MD 21250, USA \label{inst:CRESST}
\and
    Center for Space Sciences and Technology, University of Maryland, Baltimore County, 1000 Hilltop Circle, Baltimore, MD 21250, USA \label{inst:CSST}
\and
     European Space Agency (ESA), European Space Astronomy Centre (ESAC), Camino Bajo del Castillo s/n, 28692 Villanueva de la Cañada, Madrid, Spain \label{inst:esac}
\and
    Astronomy and Astrophysics Department, University of California San Diego, La Jolla CA, 92093 USA \label{inst:ucsd}
\and
    International Space Science Institute, Hallerstrasse 6, 3012 Bern, Switzerland \label{inst:issi}
\and
     Department of Physics and Astronomy, Embry-Riddle Aeronautical University, 3700 Willow Creek Road, Prescott, AZ 86301, USA \label{inst:erau}
}

\date{Received 5 December 2023 / Accepted 13 March 2024}

\bibpunct{(}{)}{;}{a}{}{,}
\newcommand{\gx}{GX\,301$-$2\xspace}
\newcommand{\nustar}{\textit{NuSTAR}\xspace}
\newcommand{\bat}{\textit{Swift}/BAT\xspace}

\begin{document}

\abstract{The High-Mass X-ray Binary (HMXB) system \gx{} is a persistent source
with a well-known variable cyclotron line centered at 35\,keV. Recently, a second cyclotron line at 50\,keV has been reported with a presumably different behavior than the 35\,keV line.}
{We investigate the presence of the newly discovered cyclotron line in the phase-averaged and phase-resolved spectra at higher luminosities than before. We further aim to determine the pulse-phase variability of both lines.}
{We analyze a \nustar{} observation of \gx covering the pre-periastron flare, where the source luminosity reached its peak of ${\sim} 4 \times 10^{37}\,\mathrm{erg}\,\mathrm{s}^{-1}$ in the 5--50\,keV range. We analyze the phase-averaged spectra in the \nustar{} energy range from 3.5--79\,keV for both the complete observation and three time segments of it. We further analyze the phase-resolved spectra and the pulse-phase variability of continuum and cyclotron line parameters.}
{We confirm that the description of the phase-averaged spectrum requires  a second absorption feature at $51.5^{+1.1}_{-1.0}$\,keV besides the established line at 35\,keV. The statistical significance of this feature in the phase-averaged spectrum is $>99.999\%$. 
We further find that the 50\,keV cyclotron line is present in three of eight phase bins.}
{Based on the results of our analysis, we confirm that the detected absorption feature is very likely to be a cyclotron line. We discuss a variety of physical scenarios which could explain the proposed anharmonicity, but also outline circumstances under which the lines are harmonically related. We further present the cyclotron line history of \gx{} and evaluate concordance among each other. We also discuss an alternative spectral model including cyclotron line emission wings.}
\maketitle

\section{Introduction}

Cyclotron Resonant Scattering Features (CRSFs), also referred to as \enquote{cyclotron lines}, are
characteristic absorption-line-like features observed in the hard X-ray spectra of a number
of accretion-powered X-ray binaries harboring a strongly magnetized
neutron star \citep[see][for a review]{Staubert2019}. The X-ray spectra are expected to be formed in structures called accretion columns near the magnetic poles of the accreting neutron star (NS). Close to the neutron star, the accreted matter is funneled towards the magnetic poles due to the strong magnetic field, $B\sim10^{12}\,\mathrm{G}$. In the plasma of the accretion column, charged particles such as electrons experience Landau quantization, where their momenta perpendicular to the magnetic field lines are quantized. Inelastic scattering of photons off these electrons can result in the formation of an absorption-line like feature at an energy corresponding to the cyclotron energy
$\sim 12$\,keV\,$B/10^{12}$\,G.

The diagnostic power of such cyclotron lines lies in the possibility to infer the magnetic field strength at the region of their formation directly from the line centroid energy. In addition, cyclotron line parameters are sensitive to the other characteristics of the accreting plasma, such as temperature, density, and the velocity of the flow. Pulse-phase-resolved studies of CRSFs have further contributed to the investigation of the accretion geomerty of neutron stars and their emission characteristics \citep[see, e.g.,][]{Maitra2017}. However, CRSFs are by no means observed in all accreting pulsars
and the exact physical conditions necessary to form an observable CRSF are not yet
well understood. Extensive observational campaigns over the last
decades \citep[e.g.,][]{Staubert2019,Pradhan21} have found variability of CRSF parameters with
rotational phase of the pulsar as well as  different types of correlations of the CRSF energy
with X-ray luminosity. Different explanations for this behavior were proposed, involving, for example,
the displacement of the line-forming region inside the accretion channel
\citep[e.g.,][]{Becker2012, Nishimura2014}, reflection from the neutron star
surface \citep[e.g.,][]{Poutanen2013}, or a Doppler shift of photons during scattering in
the accretion flow \citep{Nishimura2014, Mushtukov2015}.

In sources which exhibit multiple cyclotron lines, the relation between those is of particular interest, raising the question about their origin as to whether they are individual features from different regions or harmonic lines. The spacing of the line energies allows for probing the region of their formation and provides insights on the complex process of photon redistribution in resonant scattering events in an inhomogeneous medium.
\citet{Kendziorra1992} were the first to claim CRSFs at 54\,keV and possibly 27\,keV for Vela~X-1 \citep[see][for an extensive review]{Kretschmar2021}. Subsequent analyses confirmed the lower CRSF \citep[see, e.g.,][]{Kretschmar1997,Kreykenbohm1999}, whilst other studies \citep[see, e.g.,][]{Orlandini1998} only found the higher CRSF. The phase-resolved analysis of an \textit{RXTE} observation of Vela~X-1 by \citet{Kreykenbohm2002} detected both lines and identified them as fundamental and second harmonic, which has been confirmed by many subsequent studies \citep[see, e.g.,][]{Wang2014,Fuerst2014, Diez2022}. Another prominent X-ray source with multiple cyclotron lines is 4U\,0115+63. Its X-ray spectrum shows up to five cyclotron lines at energies of nearly integer multiples of 11\,keV \citep{Heindl1999,Ferrigno2009,Tsygankov2007}, which are interpreted as a fundamental line and its higher harmonics \citep[also, an additional fundamental line was reported at ${\sim}16\,\mathrm{keV}$,][]{Iyer2015, Liu2020}.

The focus of this work is the study of the presence and relation of the multiple cyclotron lines in \gx{} (alias 4U 1223$-$62). \gx{}, discovered by \citet{McClintock1971}, is a High Mass X-ray Binary system that contains a neutron star rotating with a period of ${\sim}700\,\mathrm{s}$ \citep{White1976} and accreting
from the stellar wind of the B hypergiant Wray\,977 \citep{Bradt1977, Kaper2006}, which is clumpy \citep{fuerst2011,Roy2023}. Under the assumption that the rotating vector model \citep[see, e.g.,][]{Radhakrishan1969} is applicable for \gx and that emission is dominated by the ordinary polarisation mode, \citet{Suleimanov2023} determine the inclination and magnetic obliquity of the source using polarimetric measurements and report values of $i_p = 135^\circ \pm 17^\circ$ and $\theta = 43^\circ \pm 12^\circ$, respectively. Using the Gaia survey \citep[Data Release 3,][]{gaiacollaboration2016GaiaMission,gaiacollaboration2022GaiaDataReleaseSummaryContent}, \citet{bailer-jones2021EstimatingDistancesParallaxesGeometricPhotogeometric} estimate a distance of \gx of $3.55^{+0.19}_{-0.17}$\,kpc.
\gx{} moves around its companion on an orbit with an eccentricity of 0.47 \citep{koh1997RapidSpinEpisodesWindFed} and an orbital period of
${\sim}41.5\,\mathrm{d}$ \citep{Watson1982,Sato1986,Leahy2002}, showing rapid orbital decay \citep{Manikantan2023}. It shows regular pre-periastron flares, where the
luminosity increases by roughly an order of magnitude compared to
other orbital phases \citep{Pravdo1995, Pravdo2001,Islam2014}. Whilst the pre-periastron flare itself is regular, the timing and luminosity of the maximum cannot be predicted precisely, which complicates the scheduling of its observation. The X-ray pulsar is a well
established CRSF source with a line centered at ${\sim}$35\,keV
\citep{Mihara1995b}. The CRSF energy is strongly variable with pulse-phase between ${\sim}$30--40\,keV \citep{Kreykenbohm2004}. Additionally, higher CRSF energies around 45--53\,keV have been reported, leading to speculation about the luminosity dependence of the CRSF \citep{LaBarbera2005,Suchy2012} and whether changes in luminosity are responsible for the observation of higher CRSF energies.

\begin{figure}
    \centering
    \resizebox{\hsize}{!}{\includegraphics{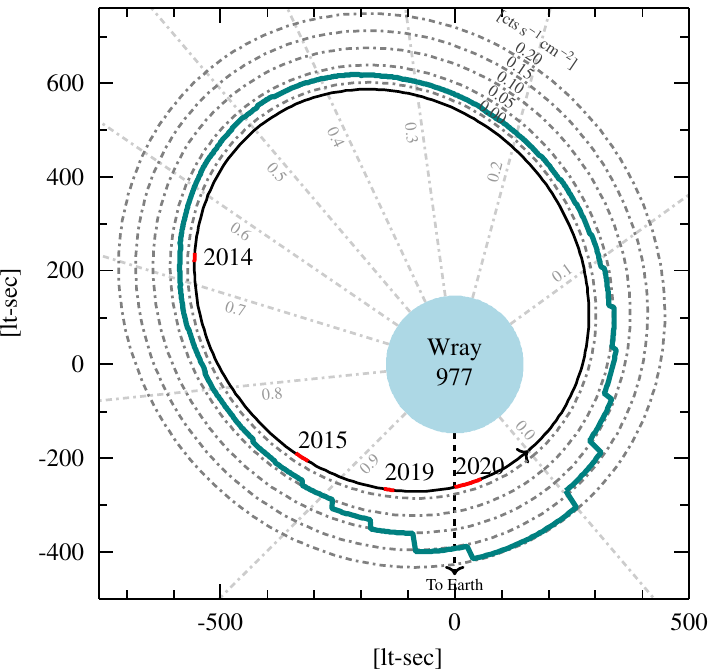}}
    \caption{Orbital geometry of the system \gx{} based on Fig.~1 of \citet{fuerst2011}. We show the neutron star orbit around its B hypergiant companion Wray 977 along with the orbital \bat \citep{Krimm2013} count rate as determined from the long-term BAT light curve in blue. We further show the orbital phases of \nustar observations of \gx and the corresponding year in red. The 2022 observation was omitted as it was performed at nearly the same orbital phase as the 2014 observation. For the orbital parameters, we use the values presented by \citet{Kaper2006}, that is $a \sin i=368.3^\circ$, $e=0.462$, $\omega=310.4^\circ$ and $R_\mathrm{Wray\: 977}=62\,R_\odot$. We use an inclination of $i=55^\circ$ \citep[within the range given by][]{Leahy2008}. Epoch folding of the long-term \bat light curve yields an orbital period of $41.474\pm0.006$\,d, which is in good agreement with the period stated by \citet{Doroshenko2010} as derived from \citet{koh1997RapidSpinEpisodesWindFed} including an orbital period derivative. We subsequently use the periastron passage $T_\mathrm{PA} = 53531.65$\,MJD from \citet{Doroshenko2010}. For the calculation of orbital phases we take the orbital period derivative $\dot P_\mathrm{orb}=-3.7\cdot 10^{-6}\,\mathrm{s}\,\mathrm{s}^{-1}$ \citep{Doroshenko2010} into account.}
    \label{fig:orbitPlot}
\end{figure}

The Nuclear Spectroscopic Telescope Array
\citep[\nustar{},][]{Harrison2013} observed the source twice between 2014--2015 at orbital phases of 0.67 and 0.87, where the luminosity was on the order of 3--4$\times10^{36}\,\mathrm{erg}\,\mathrm{s}^{-1}$, which is roughly one order of magnitude lower than during the periodic pre-periastron flares. \citet{Fuerst2018} studied the two observations along with an archival \textit{Suzaku} observation and reported two non-harmonically spaced CRSFs, one around 35\,keV and one at 50\,keV. The
existence of two separate lines was confirmed by \citet{Ding2021},
using observations of \gx by the Hard X-ray Modulation Telescope
\textit{Insight}-HXMT \citep{Zhang2020} at different orbital phases.
Their systematic analysis included data from about 40
pointed observations at different luminosities and finds a lower energy of the first CRSF of
${\sim}26$--$35\,\mathrm{keV}$ and thus higher ratio of the
centroid energies of the two CRSFs (${\approx}1.63$).
We give an overview over past \nustar observations of \gx along with the determined cyclotron line energies in Table~\ref{tab:nustarObservations}.
\begin{table*}
\renewcommand{\arraystretch}{1.3}
    \caption{Summary of \nustar observations of \gx.}
    \begin{tabular}{lrrrrrrl}
    
        \hline Date & $\varphi_\mathrm{orb}$\,\tablefootmark{a)} & $\mathcal{L}_{37}^\mathrm{NS}$\, \tablefootmark{b)} &Observation ID & Exposure [ks] & $E_\mathrm{CRSF1}$\,[keV] & $E_\mathrm{CRSF2}$\,[keV] & Reference \\ \hline
        2014-10-29 & 0.67 & 0.71 & 30001041002 & 38.2 & $34.7^{+2.1}_{-1.4}$ & $50.6^{+2.1}_{-1.7}$ & \citet{Fuerst2018} \\ 
2015-10-04 & 0.87 & 1.08 & 30101042002 & 35.7 & $34.5^{+1.6}_{-1.4}$ & $49.6^{+1.3}_{-1.2}$ & \citet{Fuerst2018} \\ 
2019-03-03 & 0.93 & 3.69 & 90501306002 & 23.0 & $36.7\pm 0.3$ & $55.1^{+0.7}_{-0.6}$ & \citet{Nabizadeh2019} \\ 
2020-12-27 & 0.97 & 5.02 & 30601013002 & 48.5 & $37.4^{+1.1}_{-1.0}$  & $51.5^{+1.1}_{-1.0}$ & This work \\ 
2022-02-02 & 0.67 & 0.55 & 30701001002 & 35.6 & $35.4^{+0.6}_{-0.2}$\,\tablefootmark{c)} & $54\pm2$\,\tablefootmark{c)} & McCulloch et al. (in prep.) \\ \hline
     \end{tabular}
    \tablefoottext{a}{Orbital phase, which is calculated in the same manner as described in the caption of Fig.~\ref{fig:orbitPlot}.}\\
    \tablefoottext{b}{Absorbed luminosity in units of $10^{37}\,\mathrm{erg}\,\mathrm{s}^{-1}$ in the energy band 5--50\,keV in the neutron star rest frame as obtained directly from the \nustar data. In order to transform the value from the observer's frame to the NS frame, the obtained values are multiplied by $(1+z)^4$, where $z=0.3$, which takes the gravitational red-shift, time dilatation and surface magnification into account.}\\
    \tablefoottext{c}{The cyclotron line parameters for the most recent observation are preliminary results and subject to further analysis.}\\
    \label{tab:nustarObservations}
\end{table*}
Furthermore, we show the orbit of the neutron star around its companion along with the orbital phases of recent \nustar observations of \gx in Fig.~\ref{fig:orbitPlot}.

\citet{Fuerst2018} argue that unlike the first CRSF, the potential second
harmonic does not show a clear phase variability.
They propose a scenario where both CRSFs originate from
different regions inside the same accretion column. In particular, one
line is supposed to form at the bottom of the column and one at
higher altitudes in a radiation-dominated shock, with the latter being
subjected to relativistic boosting and therefore showing different
pulse-phase variability. While the idea that in many accreting pulsars
only one accretion column is visible for most of the rotational period
has already been discussed in the context of pulse profile modeling
\citep{Falkner2018}, in general, CRSFs with energies different from
purely harmonic spacing have been assumed to require the superposition
of emission from two accretion columns and an asymmetric $B$-field or
mass accretion \citep[see, e.g.,][]{Liu2020}. 

For some HMXBs that show cyclotron lines, the centroid energy of the CRSF varies with the X-ray luminosity \citep[see, e.g.,][Fig. 9]{Staubert2019}. Both positive and negative correlations have been observed for different sources.
An analysis of a more recent \nustar{} observation of \gx at a luminosity of ${\sim} 0.3 \times 10^{37}\, \mathrm{erg}\,\mathrm{s}^{-1}$ \citep[in the observer's frame, derived using $d=3.55$\,kpc in the energy range of 3--79\,keV from][]{Nabizadeh2019} has resulted in higher centroid energies for both cyclotron lines compared to the observations studied by \citet{Fuerst2018}, where the source luminosity was on the order of $0.3$--$0.4 \times 10^{37}\, \mathrm{erg}\,\mathrm{s}^{-1}$ (in the observer's frame, derived using $d=3.55$\,kpc in the energy range of 5--50\,keV). \citet{Nabizadeh2019} thus conclude a positive correlation between cyclotron line energy and luminosity of \gx{}. The orbital phase-resolved analysis by \citet{Ding2021}  showed a positive luminosity-dependence of both cyclotron line energies for luminosities below $10^{37}\,\mathrm{erg}\,\mathrm{s}^{-1}$ and a negative dependence above this threshold. 

In this work, we investigate whether the newly discovered CRSF is also present during the pre-periastron flare, where the source luminosity is on the order of a few $10^{37}\,\mathrm{erg}\,\mathrm{s}^{-1}$ as well as analyze its pulse-phase dependence. We present a spectral and timing analysis of a \nustar
observation of the pre-periastron flare and compare the source
behavior to previous observations at different luminosity states. The
remainder of the paper is structured as follows: We describe the data reduction in Sect.~\ref{sec:reduction}. In Sect.~\ref{sec:pha} we describe the spectral analysis of the whole observation as well as three time-segmented spectra. In Sect.~\ref{sec:phaseresolved} we present our spectral analysis of eight phase bins, where we focus on the phase-dependence of the continuum and cyclotron line parameters. We discuss the results of the preceding sections from a physical standpoint and elaborate on their physical implications in Sect.~\ref{sec:discussion}. In Sect.~\ref{sec:conclusion} we briefly summarize the key findings of our analysis.

\section{Data Acquisition and Reduction}\label{sec:reduction}

In the following we describe the data acquisition, data reduction methods and setup for the spectral analysis.

\begin{figure}
    \centering
    \resizebox{\hsize}{!}{\includegraphics{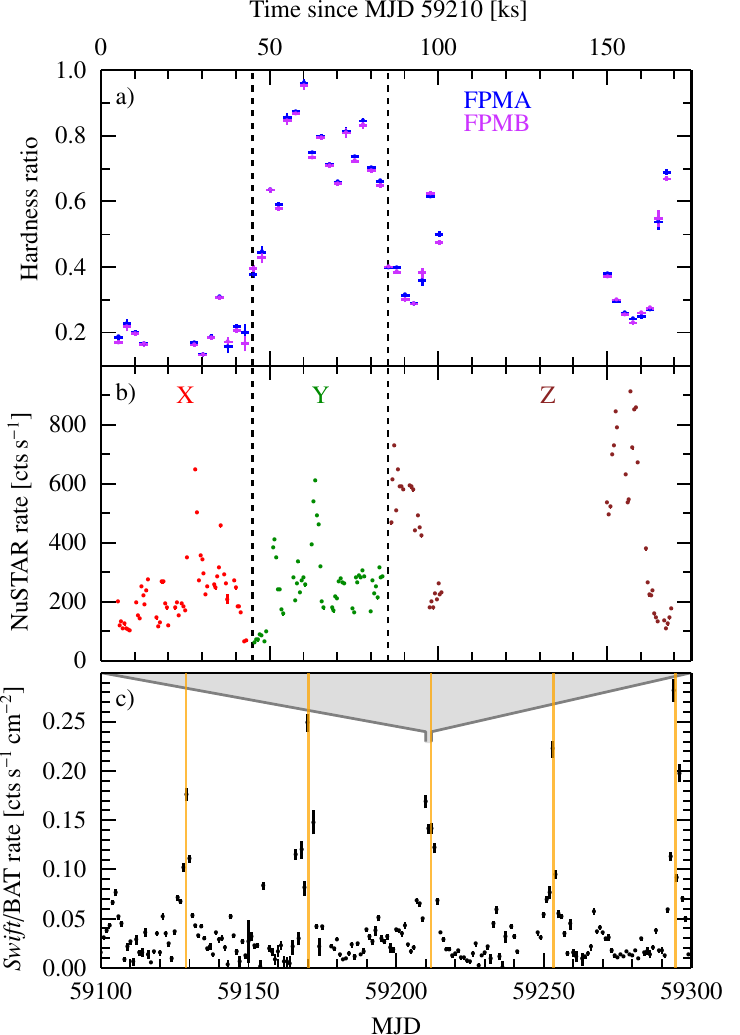}}
    \caption{\nustar observation of \gx in the context of its long-term behavior. \textbf{a}: Hardness ratio, \hbox{$(H-S)/(H+S)$}, of \nustar observation with hard energy range 10--30\,keV and soft energy range  4--10\,keV. The dashed lines indicate the division into time segments studied individually in Sect.~\ref{sec:pha_timebinned}. \textbf{b}: Light curve of \nustar observation. \textbf{c}: \bat light curve between MJD 59100 and MJD 59300 \citep{Krimm2013}. We show ${\sim}$5 orbital periods and indicate the pre-periastron flares with orange bars.}
    \label{fig:swiftbatlc}
\end{figure}

\gx was observed during one of its regular pre-periastron flares \citep{Pravdo1995} on 2020 December 27 at an orbital phase of 0.97 (as calculated from orbital parameters given in the caption of Fig.~\ref{fig:orbitPlot}) with \nustar's two nearly identical Focal Plane Modules (FPMs). We show a long-time \bat lightcurve of \gx as well as the light curve of the \nustar observation along with its hardness ratio in Fig.~\ref{fig:swiftbatlc}.
The \nustar observation captured the pre-periastron flare, but a significant fraction of the data was lost due to a telemetry issue. The resulting effective exposure time is ${\sim}48.5\,\mathrm{ks}$ per FPM.

We use \texttt{nupipeline} from HEAsoft \hbox{v.\ 6.30.1} with CALDB version 20220802 for standard reprocessing and screening, and treat both detectors individually.
We select circular source and background regions with $120 \arcsec$ radius each. The time-resolution of the extracted light curves is 1\,s.
Our analysis requires us to resolve spectral features above 50\,keV as accurately as possible. We therefore apply the rebinning criteria proposed by \citet{Kaastra2016}, which are optimized for the energy resolution of the detector rather than ensuring a certain signal-to-noise. 
We use the $\chi^2$-statistic to evaluate spectral models.

Misalignments of few tens of eV between the energy grids of the two \nustar focal plane modules with respect to the energy of the Fe\,K edge became apparent in a preliminary analysis. By introducing an additive gain shift to the energy grids of both detectors, respectively, and fitting both offset parameters $g_\mathrm{FPMA/FPMB}$, the misalignments can be corrected for.
All spectral analysis was performed with ISIS, version \mbox{1.6.2-51} \citep{Houck2000}. Unless otherwise noted, uncertainties are given at a 90\% level. Luminosities are given in the observer's frame unless stated otherwise.

\section{Phase-averaged spectroscopy} \label{sec:pha}
\gx is highly variable on all timescales from the pulse period to beyond the orbital period. In this section, we analyze the phase-averaged spectrum and its variation on the timescale of hours, i.e., the duration of the observation. Figure~\ref{fig:swiftbatlc} shows that the actual pre-periastron flare (shown with orange bars in the bottom panel) is already preceded by significant flaring activity on timescales of ks that are also associated with spectral variability \citep[see, e.g.,][]{Leahy2008,Evangelista2010}. The figure shows significant fluctuations in the hardness ratio throughout the orbital phase, aligning with the fluctuations in the absorption column density throughout the source's orbit, as reported in literature \citep{Islam2014}.  
The spectral analysis of \gx therefore requires a solid understanding of its timing properties. In Sect.~\ref{sec:wholeobs} we analyze the spectrum of the complete observation. In Sect.~\ref{sec:pha_timebinned} we divide the observation into three time segments based on changes in flux level and hardness ratio and analyze spectral changes on the timescale of hours.

\subsection{Time-averaged spectral analysis} \label{sec:wholeobs}
We conduct the spectral analysis by making use of purely empirical models. An extensive list of commonly used empirical continuum models is presented by \citet{Mueller2013} and \citet{Staubert2019}. We use a powerlaw component in combination with a Fermi-Dirac cutoff \citep[\texttt{FDcut},][]{Tanaka1986}
\begin{align}
\begin{split}
 \texttt{continuum}(E)&=\texttt{powerlaw}(E) \times \texttt{FDcut}(E)\\
  &=K \cdot E^{-\Gamma} \times \left[ 1 + \exp \left( \frac{E-E_\mathrm{cutoff}}{E_\mathrm{fold}}\right) \right]^{-1},   
\end{split}
\end{align}
where $K$ is a normalization constant, $\Gamma$ the photon index and $E_\mathrm{cutoff/fold}$ the cutoff and folding energy, respectively. The model has also been used by \citet{Kreykenbohm2004} to successfully describe the X-ray continuum of \gx during the pre-periastron flare as observed with \textit{RXTE} and more recently by \citet{Nabizadeh2019} in their analysis of a \nustar observation of \gx from 2019. \citet{Fuerst2018} on the other hand used the \texttt{NPEX} model \citep{Mihara1995b} to describe the continuum in \nustar observations from 2014 and 2015.

For the observation in focus of this publication, multiple empirical continuum models are tested in a preliminary analysis. Only a powerlaw including a \texttt{FDcut} cutoff provides a good description of the data and is therefore used throughout this spectral analysis. In general, we acknowledge the variety of continuum models that have been successfully used to describe the X-ray spectrum of \gx and emphasize that possible correlations between continuum parameters and cyclotron lines \citep[see, e.g.,][]{Mueller2013} call for caution when comparing cyclotron line parameters resulting from analyses with different continuum models.

In order to account for photoelectric absorption in the interstellar medium and by neutral material near the X-ray source, we use a partial covering model \citep[PCF, see, e.g.,][]{Ballhausen2021,Diez2022},
\begin{equation}
    \texttt{PCF}(E)=f_\mathrm{PCF} \times \texttt{tbabs}_1(E) + (1-f_\mathrm{PCF}) \times \texttt{tbabs}_2(E),
\end{equation}
where $f_\mathrm{PCF}$ is the partial covering fraction and \texttt{tbabs} is the absorption model of \citet{Wilms2000}. Our absorption model therefore has three parameters, $f_\mathrm{PCF}$ and the hydrogen column densities $N_\mathrm{H1/2}$ of the two underlying absorption components, for which we assume the cross sections by \citet[]{Verner1995} and the elemental abundances by \citet[]{Wilms2000}.
We model both CRSFs as absorption lines with Gaussian optical depth using the model \texttt{gabs} \citep[see, e.g.,][]{Staubert2019}. Its parameters are the cyclotron line energy $E_\mathrm{CRSF}$, its width  $\sigma_\mathrm{CRSF}$ and its strength $d_\mathrm{CRSF}$, from which the optical depth at $E_\mathrm{CRSF}$ can be derived as $\tau=d_\mathrm{CRSF}/(\sqrt{2 \pi} \sigma_\mathrm{CRSF})$. 
For an in-depth discussion of CRSFs and their modeling, see, e.g. \citet{Staubert2019}.
We further include components for the two prominent neutral iron emission lines Fe\,K$\alpha$ and Fe\,K$\beta$ as additive Gaussian emission lines using the model \texttt{egauss}. It is defined as
\begin{equation}
    \texttt{egauss}(E)=\frac{A}{\sqrt{2\pi}\sigma} \exp\left[-{ (E-E_0)^2 \over 2\sigma^2} \right]
\end{equation}
where $A$ is the normalization representing the area under the Gaussian curve, $E_0$ is the center of the Gaussian curve and $\sigma$ its width. The soft X-ray spectrum of \gx{} was studied in detail by \citet{Watanabe2003} using \textit{Chandra} High-Energy Transmission Grating (HETG) observations, which show that the widths of the iron emission lines are smaller than the energy resolution of \nustar{} \citep[${\sim}400$\,eV at 6.4\,keV, see][]{Harrison2013}, for which reason we do not expect to be able to constrain them. We therefore fix the line widths to 1\,eV. We furthermore fix both line energies to their theoretical values as given in the X-Ray Data Booklet\footnote{See \url{https://xdb.lbl.gov/}} \citep[based on][]{Bearden1967,Krause1979}, that is $E_{\mathrm{Fe\,K}\alpha}=6.397$\,keV and $E_{\mathrm{Fe\,K}\beta}=7.058$\,keV. The centroid energy of the Fe\,K$\alpha$ line was determined as the average of the two Fe energy levels contributing to the observed emission line at 6.391\,keV and 6.404\,keV. The study of the fluorescence lines is outside the scope of this paper, we therefore refer to \citet{fuerst2011} for a detailed study of the fluorescence lines in the X-ray spectrum of \gx.

Our preliminary analysis shows the presence of an excess at the lower end of the \nustar energy range at 3.5--5.0\,keV, which cannot be described correctly with the continuum model. We therefore include a blackbody component \texttt{bbody} to account for the soft excess. Using the above model provides no satisfactory description of the soft X-ray spectrum at energies of 5--7\,keV. We therefore include an additional Gaussian line (hereafter denoted by index SE for soft emission), with centroid energy of ${\sim}$6\,keV and width below 1\,keV. An extensive justification for this component is given in Appendix~\ref{sec:AppendixB}. We note that this added emission component is far narrower than the blackbody component and cannot be compensated for by varying blackbody temperature and normalization.

To account for detector flux calibration uncertainties a multiplicative factor $C_\mathrm{FPMB}$ for the FPMB spectrum is introduced. This parameter is expected to be well within ${\sim}$3\% of unity \citep[see, e.g.,][]{Madsen2022}. To account for a shift in the energy grids of both detectors we include additive gain shift components ($g_\mathrm{FPMA}$ and $g_\mathrm{FPMB}$), which shift the energy grids according to their value. The total model can therefore be expressed as
\begin{align}\label{eq:baseline}
\begin{split}
    f(E)=&\texttt{detconst} \times \bigl[ \bigl(\texttt{continuum}(E) + \texttt{bbody}(E)\bigr) \times \\
    &\texttt{PCF}(E) \times \texttt{gabs}_1(E) \times \texttt{gabs}_2(E) + \\
    &\texttt{egauss}_\alpha(E) + \texttt{egauss}_\beta(E) + \texttt{egauss}_\mathrm{SE}(E)\bigr]~.
\end{split}
\end{align}
The two iron lines and the additional Gaussian line are not subject to absorption in our spectral model as defined in Eq.~\eqref{eq:baseline}. The use of either unabsorbed or absorbed emission lines constitutes a simplification of the expected conditions, where the reprocessing of radiation within the ambient medium gives rise to the observed emission as well as absorption features simultaneously. A preliminary analysis shows that both fit statistic as well as line shape do not differ from a model in which the emission lines are also absorbed, we therefore use a model in which the lines are not absorbed. 

\begin{figure}
    \centering
    \resizebox{\hsize}{!}{\includegraphics{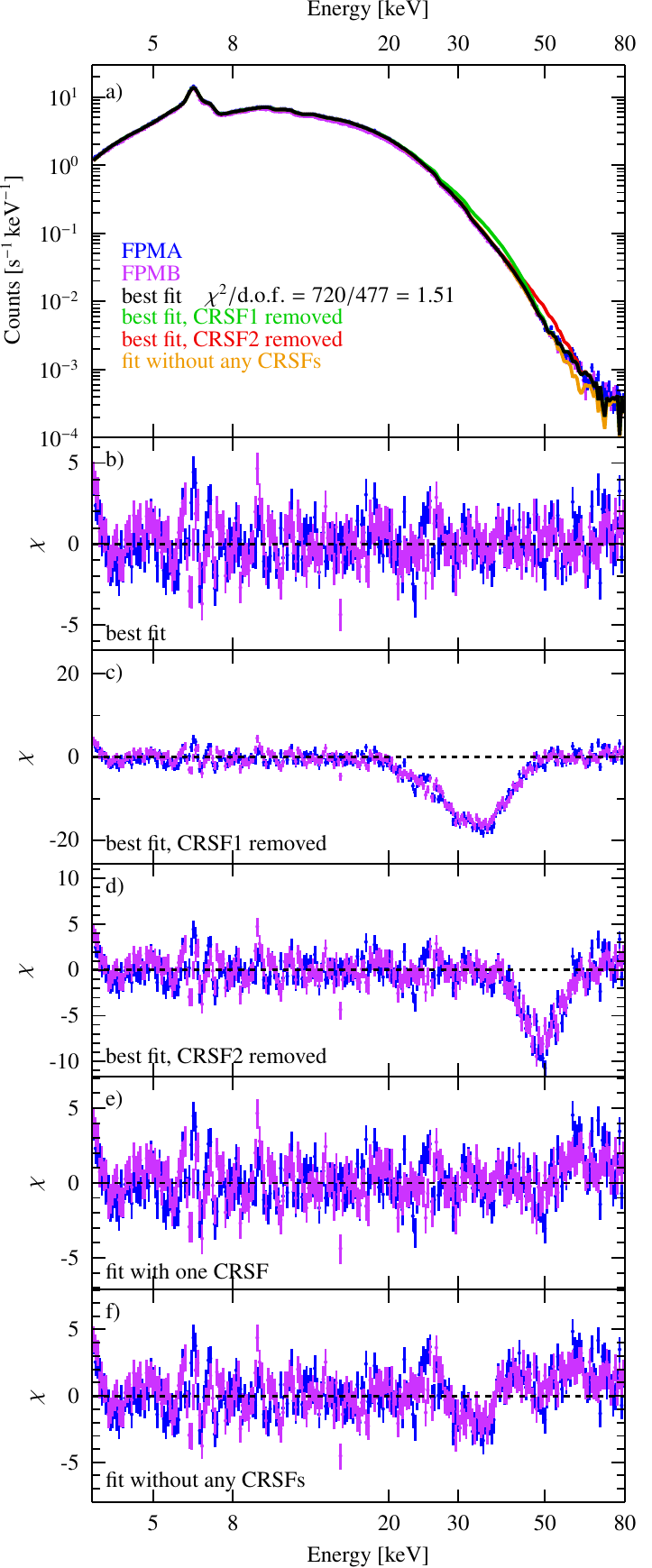}}
    \caption{Continuum modeling of \gx. \textbf{a} spectra of the two \nustar{} FPMs in blue and purple with best-fit model shown in black. Models with both CRSFs removed in turn and model without CRSFs shown in green, red and orange. \textbf{b}: $\chi$ residuals of best fit. \textbf{c} and \textbf{d}: Residuals of best-fit with both CRSFs removed in succession. \textbf{e}: Best-fit of alternative model with only one CRSF. \textbf{f}: Best-fit of alternative model without CRSFs. The spectra have been rebinned for plotting purposes.}
    \label{fig:phaseAveraged}
\end{figure}

In the remainder of this paper, the index 1 and term \enquote{lower line} denote the CRSF first claimed by \citet{Mihara1995b} around 35\,keV; the index 2 and term \enquote{higher line} denote the CRSF discovered by \citet{Fuerst2018} around 50\,keV. Furthermore, unless otherwise stated, parameters by \citet{Fuerst2018} used for comparison to our findings are results of their \texttt{NPEX} fit to the \nustar observation from 2015 (see their Table~2).

Our model with two cyclotron lines provides the most acceptable fit to the data among the tested models, with a goodness of fit statistics of $\chi^2 / \mathrm{d.o.f.}=720/477=1.51$. We show the observed spectrum along with the best-fit model in the upper panel of Fig.~\ref{fig:phaseAveraged}. All free fit parameters are given in Table~\ref{tab:TRS_parameters}. We will call this model the Whole Observation Fit (WOF) in the remainder of this paper. We further show $\chi$ residuals in the second panel as well as residuals of the fit with both cyclotron lines removed in succession in order to illustrate their impact on the spectral fit.
The residuals of the best-fit model do not deviate significantly from the data and do not show a systematic structure, but rather stochastic variability. The only exception is an excess  below 4\,keV, which we attribute to the use of a simplified absorption model; this deviation does however not affect the analysis of the CRSFs. 

The energy of the lower cyclotron line is $E_\mathrm{CRSF1}=37.4^{+1.1}_{-1.0}$\,keV, which is roughly 3\,keV higher than the value reported by \citet{Fuerst2018} for observations at different orbital phase and luminosity. The width of the line is found to be $\sigma_\mathrm{CRSF1}=6.6^{+1.3}_{-0.9}$\,keV, with a strength $d_\mathrm{CRSF1}=5.5^{+2.4}_{-1.5}$\,keV, corresponding to an optical depth $\tau_1=0.33$ at the centroid energy.
The second cyclotron line has a line energy of $E_\mathrm{CRSF2}=51.5^{+1.1}_{-1.0}$\,keV. This value is  ${\sim}2$\,keV higher than the values reported by \citet{Fuerst2018}. We further report a width of $\sigma_\mathrm{CRSF2}=5.1^{+1.0}_{-0.8}$\,keV and a strength of $d_\mathrm{CRSF2}=7.7^{+3.2}_{-2.2}$\,keV, i.e., an optical depth $\tau_2=0.60$ at $E_\mathrm{CRSF2}$.
It is evident that the lower CRSF is wider than the high-energy line. Since thermal broadening is proportional to the line energy, other broadening mechanisms are required to explain the observed behavior.

From an alternative fit model in which the Gaussian emission lines are subject to the partial covering absorption component, we determine the equivalent width of the absorbed Fe\,K$\alpha$ emission line to be $\mathrm{EW}_{\mathrm{K}\alpha}=401\,$eV. This value is roughly a factor of 3 higher than that obtained by \citet{Fuerst2018}, who report a value of $\mathrm{EW}_{\mathrm{K}\alpha}=134\,$eV for their second \nustar observation, whose luminosity is lower by a factor of ${\sim}$8 compared to the observation studied here.
During our spectral analysis we further tested models containing one and zero CRSFs. The residuals of these models are shown in Fig.~\ref{fig:phaseAveraged}e and~f, and their CRSF parameters are given in Table~\ref{tab:tabCRSF}. Both the residuals and fit statistic of the fit without CRSFs indicates that the model lacks a required component in the range 30--50\,keV. The model with one CRSF contains a comparably narrow and shallow CRSF at 35\,keV. The residuals indicate the presence of a second possibly also narrow CRSF at energies ${\sim}$50\,keV. The inclusion of such a component however changes the parameters of the lower CRSF in an unexpected way: its centroid energy is shifted by 3\,keV toward higher energies whilst nearly doubling in width and increasing sixfold in strength, which implies an increase in optical depth by factor 4. With a width of 5.1\,keV, the second CRSF is also wider than expected from the residuals in Fig.~\ref{fig:phaseAveraged}, panel~e. Investigation of the parameter space shows that no statistically favorable set of parameters that preserves the shallow and narrow nature of the lower cyclotron line exists. While the two features show very minor overlap at most, they nevertheless influence each other indirectly via changes in the overall continuum shape, which can lead to the observed unexpected change in CRSF parameters.

\begin{table}
    \caption{CRSF parameters and goodness of fit of models including zero, one and two CRSF components. The residuals of the respective models are shown in Fig.~\ref{fig:phaseAveraged}, panels f, e and~b.}
    \renewcommand{\arraystretch}{1.3}
\begin{tabular}{lrrr}
        \hline Fit with & 0 CRSFs & 1 CRSF & 2 CRSFs \\ \hline
         $E_\mathrm{CRSF1}$ [keV] & -- & $34.5\pm0.5$ & $37.4^{+1.1}_{-1.0}$ \\
$\sigma_\mathrm{CRSF1}$ [keV] & -- & $3.7^{+0.6}_{-0.5}$ & $6.6^{+1.3}_{-0.9}$ \\
$d_\mathrm{CRSF1}$ [keV] & -- & $0.86^{+0.19}_{-0.15}$ & $5.5^{+2.4}_{-1.5}$ \\
$E_\mathrm{CRSF2}$ [keV] & -- & -- & $51.5^{+1.1}_{-1.0}$ \\
$\sigma_\mathrm{CRSF2}$ [keV] & -- & -- & $5.1^{+1.0}_{-0.8}$ \\
$d_\mathrm{CRSF2}$ [keV] & -- & -- & $7.7^{+3.2}_{-2.2}$\\ \hline 
$\chi^2/\mathrm{d.o.f.}$ & $1065/483$ & $820/480$ & $720/477$\\ 
$\chi^2_\mathrm{red}$ & $2.20$ & $1.71$ & $1.51$  \\ \hline
    \end{tabular}
    
    \label{tab:tabCRSF}
\end{table}

We further note that the fitted gain offsets are $-32.2$\,eV and $-61.2$\,eV for the two focal plane modules, which corresponds to ${\sim}$0.75 and ${\sim}$1.5 \nustar energy bins, respectively. \citet{Grefenstette2022} quote expected gain offsets of ${\approx}$40\,eV, which indicates that the found offsets are within the expected range. This is also consistent with \citet{Diez2023}, who find an offset between \nustar and \textit{XMM}-Newton of $-87$\,eV and \citet{Ballhausen2020}, who obtained gain offsets of $-25$\,eV and $-67$\,eV for the two \nustar{} FPMs, respectively. We therefore conclude that the obtained gain offsets are within the expected range.

We stress that while the residuals alone are a sufficiently strong argument for the inclusion of a second CRSF, we back up our finding by use of statistical means. The improvement in $\chi^2$ statistic by including a second CRSF with three additional free parameters is found to be $\Delta \chi^2=100$. We aim to show that the statistic improvement found above stems from the correct modeling of an actual observed absorption feature and is not merely a consequence of adding degrees of freedom to the fit model until it describes the spectrum sufficiently well.
We therefore simulate spectra in order to show the statistical significance of the CRSF detection.

We simulate 100\,000 observations of \gx{} based on the spectral model described in Eq.~\eqref{eq:baseline} with only CRSF1. The parameters are drawn from Gaussian distributions  centered around the best-fit value as given in the left-most column of Table~\ref{tab:TRS_parameters} and variance derived from the parameter uncertainty. We therefore make use of our best-fit model but remove CRSF2 manually. Alternatively, one could use the parameters of a model that contains only one CRSF ab initio. The statistical results of the two approaches may differ slightly as the parameters of CRSF1 change upon the inclusion of CRSF2. We argue however that the best description of the phase-averaged spectrum  is obtained with CRSF1 as stated in the last column of Table~\ref{tab:tabCRSF}, i.e., in a fit with both CRSFs; we therefore base our statistical test on the best-fit including both lines, where CRSF2 is removed manually. The spectra are simulated based on Poisson statistics. 

We fit both models with one and two CRSFs respectively to the data.
In the 100\,000 simulated spectra and the corresponding fit models, we find no simulation showing an improvement of more than $\Delta \chi^2 =100$ in fit statistic upon the inclusion of CRSF2;
we therefore detect the second CRSF with a statistical significance of $>99.999\%$. 

\citet{Fuerst2018} carried out a similar test and found one false positive out of 10\,000 simulated and tested spectra, which corresponds to a false positive rate of $\leq 0.01\%$.
It is noteworthy that the statistical test of the 2020 \nustar observation shows a much higher detection confidence, which is most likely due to the fact that the observation at hand has a higher luminosity of factor ${\sim}$8.

We lastly comment on the present statistical test, which is used both in this work as well as by \citet{Fuerst2018}. We assume that parameters are Gaussian-distributed, which is not necessarily the case. Moreover, drawing parameters independently from each other implies the assumption that they are uncorrelated, which is generally not justified for such a phenomenological model. As a result of this, we most likely sample a much larger parameter space than intended, which also contains spectra which do not agree with out best-fit model within uncertainty. Therefore, our statistical test should be understood as a lower limit of the significance of the presence of CRSF2 in the phase-averaged spectrum.

Apart from the solution presented here with two CRSFs, another solution with comparable fit statistic exists, in which $\sigma_\mathrm{CRSF1}$ exceeds 10\,keV. Cyclotron lines are expected to be broadened by a variety of causes: thermal motion of electrons and inhomogenities in the conditions of the line-forming region such as magnetic field strength, plasma temperature and density. Lastly, the viewing angle on the accretion column is also expected to impact the observed line shape in the phase-averaged spectrum \citep[see, e.g.,][]{Schwarm2017Thesis}, which introduces a pulse-phase variability as the viewing angle changes over the pulse period. All of the above can lead to wider CRSF profiles and potentially complex line shapes. Here, however, we argue that the \enquote{narrow-CRSF} solution for the phase-averaged spectrum is more physically motivated for the following reasons:

\begin{enumerate}
    \item The wide CRSF removes ${\sim}$5\% of the total counts, whereas the narrow CRSF in our best-fit model removes  ${\sim}$1\%. Although physically plausible cases of exceptionally influential CRSFs have been found before \citep{Nakajima2010,Malacaria2023}, \citet[Table~A.5]{Staubert2019} show that CRSFs with $\sigma>10\,$keV are more the exception than the rule. Furthermore, the table shows that such wide CRSFs are predominantly found for high-energy CRSFs with line energies $E_\mathrm{CRSF}\gtrsim 50$\,keV.
    \item The residuals shown in Fig.~\ref{fig:phaseAveraged}f do not support the claim of a cyclotron line width $>$10\,keV as the absorption feature at ${\sim}$35\,keV is at most ${\sim}$8\,keV wide.
    \item The existence of the solution with a broad CRSF parameters can be explained by correlations between continuum and CRSF parameters \citep[see, e.g.,][]{Mueller2013}, where the CRSF models a part of the continuum and no longer describes the absorption feature caused by cyclotron resonant scattering.
\end{enumerate}
We therefore continue with the narrow-CRSF solution described and shown above as the basis of physical interpretation of the phase-averaged spectrum.

\subsection{Time-segmented spectral analysis} \label{sec:pha_timebinned}
The upper panel of Fig.~\ref{fig:swiftbatlc} shows a significant temporal change in the hardness ratio of the 4--10\,keV and 10--30\,keV bands. This variation could be due to a change of the continuum shape, which in turn could influence the modeling of the CRSFs. In order to better understand the temporal variation of the spectrum that precedes the pre-periastron flare and the change in absorption during the flare, we divide the observation in three time segments marked as X,Y and Z (\textbf{T}ime\textbf{S}egment X, TS~Y, and TS~Z). The segmentation is carried out in consideration of changes in both the hardness ratio and flux level of the source as indicated by the vertical line in the upper and middle panel of Fig.~\ref{fig:swiftbatlc}. Alternatively to such a temporal division, one can argue that a division only with regard to hardness ratio is more appropriate, but such an approach lacks the ability to gain an understanding of the interesting temporal evolution of the source at the onset of the pre-periastron flare. Time segment Z coincides with the flaring event, where the source reaches its peak luminosity \citep[see, e.g.,][their Fig.~1, or the lower panel of Fig.~\ref{fig:swiftbatlc}]{Kreykenbohm2004}.

\begin{figure}
    \centering
    \resizebox{\hsize}{!}{\includegraphics{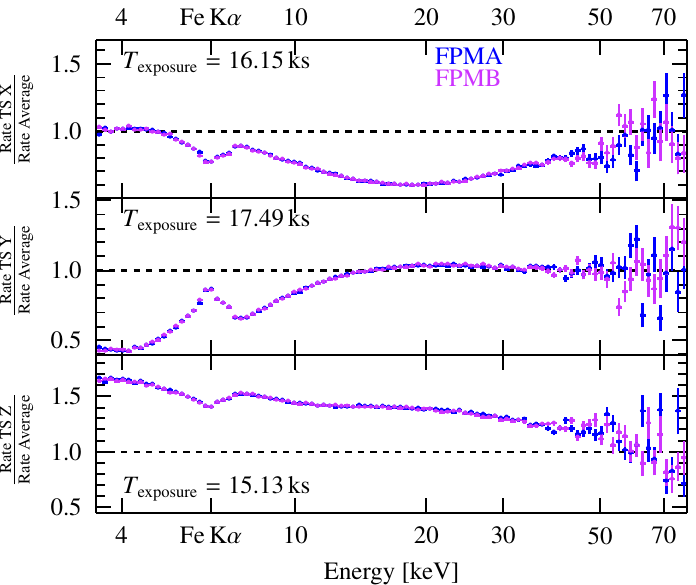}}
    \caption{Ratios between spectra of the three time segments versus the spectrum of the whole observation. The dashed line indicates the behavior expected in the case of no spectral change.}
    \label{fig:TRS_ratios}
\end{figure}

In order to identify potential changes in the spectral components between the three time segments, we start by inspecting the ratio between the time segment spectra and that of the whole observation, which are shown in Fig.~\ref{fig:TRS_ratios}. Deviations from a horizontal line indicate possible spectral changes during the time segment with respect to the full observation. Most notably, the three ratios do not match in shape or trend over the whole spectral range. Above 10\,keV, especially TS~X and TS~Y show significant curvatures of concave and convex nature with respect to the expected horizontal line, respectively. The most profound differences however are present in the energy band up to 10\,keV, most notably around the prominent iron line at 6.4\,keV. This behavior is a potential indicator of a change in absorption during the observation.

\begin{figure}
    \centering
    \resizebox{\hsize}{!}{\includegraphics{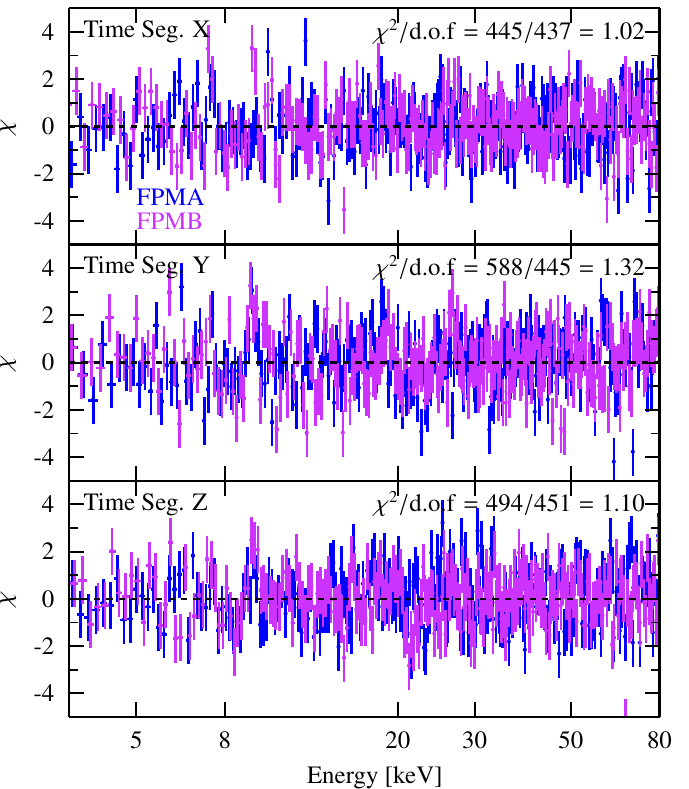}}
    \caption{Residuals of best-fits to time-segmented spectra.}
    \label{fig:TRS_residuals}
\end{figure}

To assess the spectral changes among the time-segmented spectra which are indicated by the ratios shown in Fig.~\ref{fig:TRS_ratios}, we apply the continuum model (Eq.~\ref{eq:baseline}) to the individual spectra extracted for the three time segments. The $\chi$ residuals of the fits are shown in Fig.~\ref{fig:TRS_residuals}, the fit parameters are given in Table~\ref{tab:TRS_parameters}.

Apart from few outliers at ${\sim}$6\,keV the residuals of the fits to the time-segmented spectra show no systematic deviation from the data. We find that the absorption model parameters change during TS~Y as $N_\mathrm{H,1}$ increases by a factor of 1.7 whilst $N_\mathrm{H,2}$ decreases by 25\%. The partial covering fraction decreases over the whole observation from $0.64$ to~$0.46$. The continuum parameters such as photon index, cutoff and folding energy also differ between the time segments, which is expected due to the observed curvatures in Fig.~\ref{fig:TRS_ratios}. The blackbody temperature increases gradually over the observation. For TS~X and TS~Y, the upper limit of its contribution to the spectrum cannot be constrained; for TS~Y we find a relative norm that exceeds the value obtained for the whole observation by a factor of ${\sim}$60, which most likely massively overestimates the actual contribution of the blackbody component, as we expect its relative contribution with respect to the continuum to be well below a factor of 10. The contribution of both iron lines increases over the duration of the observation, the normalization of the Gaussian soft component increases from TS~X to TS~Y and remains constant within uncertainty from TS~Y to TS~Z.

We now turn to the parameters in the focus of this study, those of the CRSFs. The line energy of the lower line remains constant within uncertainty intervals and is consistent with the value determined for the whole observation in all time segments. The line widths and strengths during TS~X and TS~Z are in agreement with those width obtained for the whole observation, for TS~Y the width increases to 10\,keV and its strength increases to 14\,keV. For CRSF2, we also find that the line energies are consistent with the ones obtained in Sect.~\ref{sec:wholeobs}. Both the width and strengths found in the time-segmented spectra are also consistent with the WOF.  We note that due to the reduced number of counts in the segmented spectra the CRSF parameters are generally less well-constrained as in the fit of the whole observation.

In conclusion, we find that both the continuum shape and absorption are variable on the timescales of the observation. The cyclotron line parameters however vary only negligibly and are mostly consistent with the WOF values as shown by the above time-segmented analysis. The above claim is also supported by a preliminary analysis of the whole observation in the hard X-ray band from 10--79\,keV, which poses the advantage of being mostly unaffected by changes in absorption; it further confirms that the excess below 4\,keV in Fig.~\ref{fig:phaseAveraged}b does not affect the analysis of the cyclotron lines. We therefore conclude that it is justified to conduct the subsequent phase-resolved analysis for the whole observation together instead of a time-segmented manner as in Sect.~\ref{sec:pha_timebinned}.

\begin{table*}
\renewcommand{\arraystretch}{1.3}
    \caption{Fit parameters of most relevant fits to the phase-averaged spectrum. We show the parameters of the fit to the whole observation and the three time segments.}
\begin{tabular}{lrrrr}
         \hline Parameter & WOF & TS~X & TS~Y & TS~Z \\ \hline
         $N_\mathrm{H,1} \: [10^{22} \, \mathrm{cm}^{-2}]$ & $30.0\pm1.3$ & $24.8^{+1.6}_{-5.4}$ & $41\pm4$ & $25.8^{+2.0}_{-1.6}$\\ 
$N_\mathrm{H,2} \: [10^{24} \, \mathrm{cm}^{-2}]$ & $2.94^{+0.28}_{-0.34}$ & $2.59^{+0.30}_{-0.74}$ & $1.93^{+0.32}_{-0.26}$ & $2.69^{+0.22}_{-0.19}$\\ 
$f_\mathrm{pcf}$ & $0.514^{+0.028}_{-0.029}$ & $0.64^{+0.14}_{-0.05}$ & $0.543^{+0.034}_{-0.030}$ & $0.46^{+0.04}_{-0.05}$\\ 
$F_1\,[\mathrm{keV}\,\mathrm{s}^{-1}\,\mathrm{cm}^{-2}]$\,\tablefootmark{a)} & $11.7\pm0.6$ & $7.0^{+0.5}_{-0.7}$ & $10.6^{+0.6}_{-0.4}$ & $16.8^{+1.6}_{-0.9}$\\ 
$\Gamma$ & $0.28^{+0.15}_{-0.16}$ & $0.94^{+0.08}_{-0.19}$ & $0.35\pm0.09$ & $0.59^{+0.13}_{-0.12}$\\ 
$E_\mathrm{fold}$ [keV] & $6.50^{+0.20}_{-0.18}$ & $7.0^{+0.5}_{-1.7}$ & $6.1^{+0.4}_{-0.7}$ & $6.39^{+0.38}_{-0.29}$\\ 
$E_\mathrm{cutoff}$ [keV] & $14.2^{+2.5}_{-2.6}$ & $22.2^{+10.7}_{-2.2}$ & $20.1^{+5.1}_{-2.9}$ & $16.8^{+2.1}_{-1.7}$\\ 
$N_\mathrm{BB}$\,\tablefootmark{b)} & $1.19^{+0.30}_{-0.24}$ & $8^{+93}_{-5}$ & $67^{+34}_{-29}$ & $1.7^{+1.5}_{-0.8}$\\ 
$T_\mathrm{BB}$ [keV] & $0.439^{+0.028}_{-0.033}$ & $0.25^{+0.04}_{-0.06}$ & $0.264^{+0.022}_{-0.016}$ & $0.341^{+0.050}_{-0.030}$\\ 
$A_{\mathrm{Fe\,K} \alpha}$ [$10^{-3}$\,ph\,s$^{-1}$\,cm$^{-2}$] & $9.78\pm0.12$ & $6.2^{+0.7}_{-0.4}$ & $10.32\pm0.19$ & $12.4^{+0.4}_{-0.6}$\\ 
$A_{\mathrm{Fe\,K} \beta}$ [$10^{-3}$\,ph\,s$^{-1}$\,cm$^{-2}$] & $0.84\pm0.10$ & $0.62^{+0.17}_{-0.15}$ & $0.96\pm0.14$ & $1.41^{+0.20}_{-0.29}$\\ 
$A_\mathrm{SE}$ [$10^{-3}$\,ph\,s$^{-1}$\,cm$^{-2}$] & $8.0^{+1.0}_{-1.2}$ & $2.0^{+2.9}_{-1.0}$ & $6.4^{+0.7}_{-0.8}$ & $5.8^{+1.9}_{-2.3}$\\ 
$E_\mathrm{SE}$ [keV] & $6.289\pm0.021$ & $6.16^{+0.10}_{-0.13}$ & $6.310\pm0.027$ & $6.28^{+0.05}_{-0.06}$\\ 
$\sigma_\mathrm{SE}$ [keV] & $0.61^{+0.05}_{-0.06}$ & $0.23^{+0.20}_{-0.14}$ & $0.51^{+0.05}_{-0.04}$ & $0.42^{+0.13}_{-0.17}$\\ 
$E_\mathrm{CRSF1}$ [keV] & $37.4^{+1.1}_{-1.0}$ & $37.9^{+1.8}_{-1.5}$ & $38.0^{+1.4}_{-2.2}$ & $37.3^{+1.7}_{-1.5}$\\ 
$\sigma_\mathrm{CRSF1}$ [keV] & $6.6^{+1.3}_{-0.9}$ & $7.3^{+4.2}_{-1.9}$ & $10.0^{+3.1}_{-2.5}$ & $6.9^{+2.2}_{-1.5}$\\ 
$d_\mathrm{CRSF1}$ [keV] & $5.5^{+2.4}_{-1.5}$ & $5.2^{+24.9}_{-2.6}$ & $14^{+17}_{-8}$ & $5.8^{+5.3}_{-2.7}$\\ 
$E_\mathrm{CRSF2}$ [keV] & $51.5^{+1.1}_{-1.0}$ & $50.8^{+1.3}_{-1.0}$ & $52.5^{+2.1}_{-1.6}$ & $51.3^{+2.3}_{-2.4}$\\ 
$\sigma_\mathrm{CRSF2}$ [keV] & $5.1^{+1.0}_{-0.8}$ & $3.6^{+1.5}_{-1.0}$ & $5.0^{+1.8}_{-1.2}$ & $4.9^{+1.9}_{-1.6}$\\ 
$d_\mathrm{CRSF2}$ [keV] & $7.7^{+3.2}_{-2.2}$ & $5.2^{+3.8}_{-2.0}$ & $6.9^{+6.4}_{-3.0}$ & $6^{+6}_{-4}$\\ 
$C_\mathrm{FPMB}$ & $1.0035\pm0.0011$ & $1.0006\pm0.0023$ & $1.0030\pm0.0020$ & $1.0036\pm0.0017$\\ 
$g_\mathrm{FPMA}$\, [eV] & $-32.2\pm2.0$ & $-18^{+8}_{-14}$ & $-32.1\pm2.9$ & $-28^{+7}_{-5}$\\ 
$g_\mathrm{FPMB}$\, [eV] & $-61.2\pm2.0$ & $-48^{+8}_{-14}$ & $-59.1\pm2.9$ & $-59^{+7}_{-5}$\\ \hline 
$\mathcal{L}_{37}$\,\tablefootmark{c)} & $2.83^{+0.14}_{-0.13}$ & $1.69^{+0.11}_{-0.17}$ & $2.55^{+0.13}_{-0.10}$ & $4.06^{+0.38}_{-0.21}$ \\ 
$\chi^2/\mathrm{d.o.f.}$ & $720/477$ & $445/437$ & $588/445$ & $494/451$\\ 
$\chi^2_\mathrm{red}$ & $1.51$ & $1.02$ & $1.32$ & $1.10$\\ \hline 
     \end{tabular}\\
    \tablefoottext{a}{Unabsorbed flux of the continuum and blackbody in the energy range of 5--50\,keV.}\\
    \tablefoottext{b}{Relative norm of the blackbody component with respect to the powerlaw component.}\\
    \tablefoottext{c}{Luminosity in units of $10^{37}\mathrm{erg}\,\mathrm{s}^{-1}$ in the energy range of 5--50\,keV and observer's frame.}
    \label{tab:TRS_parameters}
\end{table*}

\section{Phase-resolved spectroscopy}\label{sec:phaseresolved}
\begin{figure}
    \centering
    \resizebox{\hsize}{!}{\includegraphics[width=\linewidth]{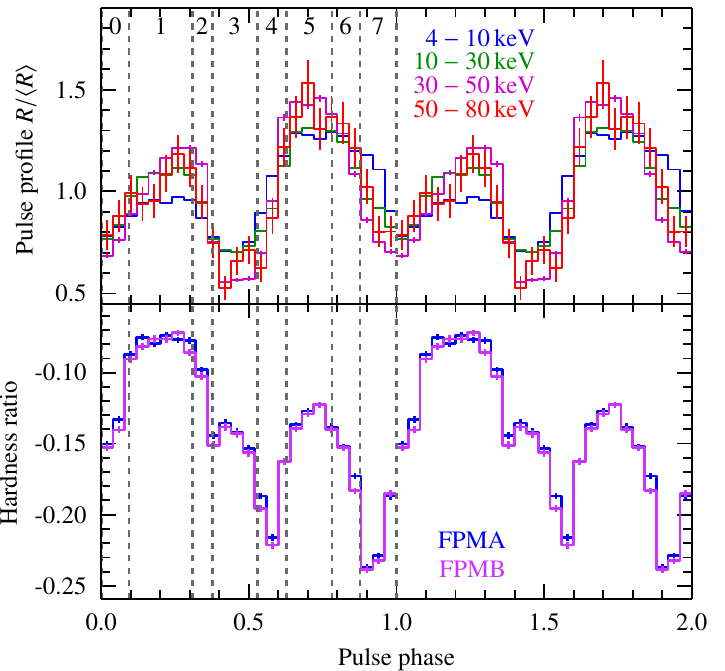}}
    \caption{Behavior of \gx over the pulse-phase. Top panel: Pulse profiles for different energy bands. Bottom panel: Hardness ratio \mbox{$(H-S)/(H+S)$} with soft energy range 4--10\,keV and hard energy range 10--30\,keV.}
    \label{fig:prs_ppHR}
\end{figure}

After analyzing the phase-averaged spectrum of \gx, we now turn to the analysis of the phase-resolved spectrum. Using epoch folding \citep[see, e.g.,][]{Leahy1983}, we determine the pulse period of the neutron star to be $P=671.62\pm0.05$\,s based on \nustar light curves with 1\,s binning. We show both the pulse profile for different energy bands as well as the hardness ratio in Fig.~\ref{fig:prs_ppHR}. The pulse profile shows two peaks. We define the phase zero as the minimum between the main and weaker pulse. The main peak is at a phase of $\varphi \sim 0.6$--$0.9$ and a weaker peak at $\varphi \sim 0.1$--$0.3$. A similar pulse profile structure was also obtained by \citet[their Fig.~3]{AlonsoHernandez2022}, who analyzed the 0.5--78\,keV pulse profile of \gx{}.

\begin{figure}
    \centering
    \resizebox{\hsize}{!}
    {\includegraphics{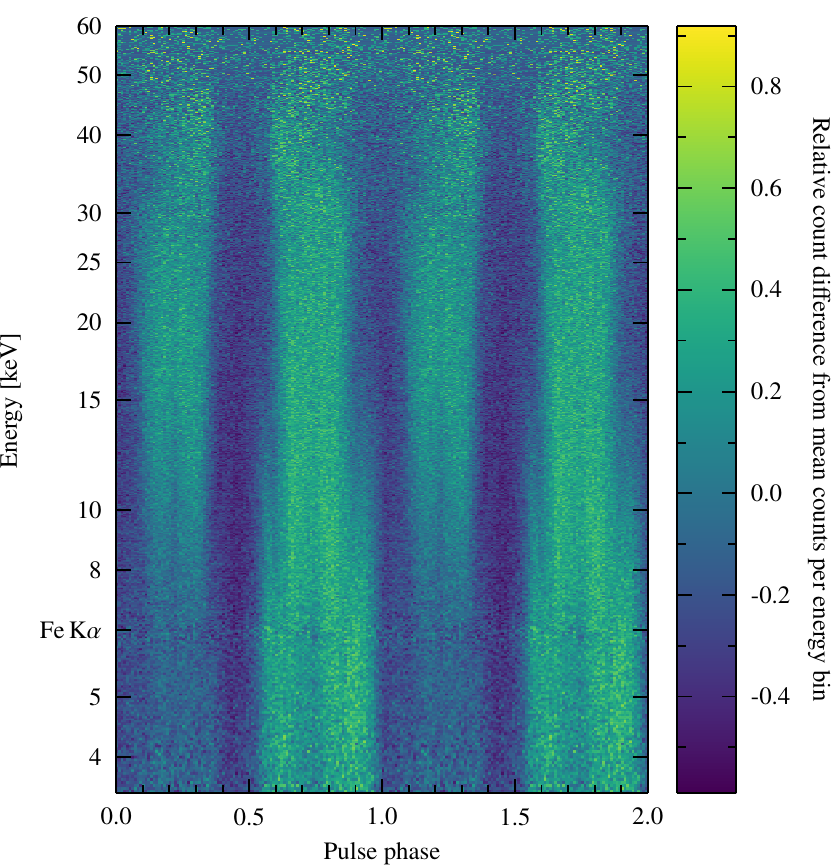}}
    \caption{Phase energy map for energies between 3.5--60\,keV. For each energy bin we show $(N- \langle N \rangle)/(N_\mathrm{max}-N_\mathrm{min})$.}
    \label{fig:prs_phEmap}
\end{figure}

\begin{figure}
    \centering
    \resizebox{\hsize}{!}{\includegraphics{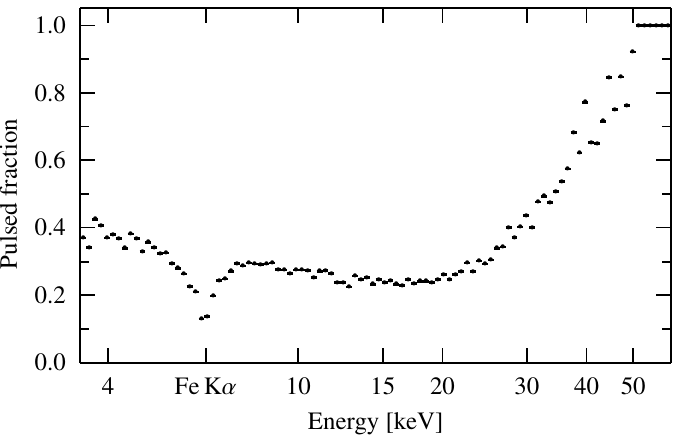}}
    \caption{Pulsed fraction as a function of energy. We define the pulsed fraction as $(N_\mathrm{max}-N_\mathrm{min})/(N_\mathrm{max}+N_\mathrm{min})$, where $N_\mathrm{max/min}$ are the maximum and minimum rates of the pulse profile in a given energy bin.}
    \label{fig:pulsedfrac}
\end{figure}

To further illustrate \gx{}'s behavior over pulse-phase, we show a phase energy map, where count rates are shown as a function of both pulse-phase and energy, in Fig.~\ref{fig:prs_phEmap}.
The peak at $\varphi \sim 0.2$ is most prominent at energies exceeding 10\,keV and less pronounced at energies below ${\sim}$6\,keV. The main peak at $\varphi \sim 0.7$, on the other hand, is also present in the soft X-ray band. Both peaks show a splitting into two components towards lower energies. At energies above 35\,keV a phase shift of the peaks is present. To further investigate the pulsations and especially whether the Fe\,K$\alpha$ shows weaker pulsations than other soft components, we show the pulsed fraction as a function of energy in Fig.~\ref{fig:pulsedfrac}.
It is evident that the pulsed fraction is highly variable with energy.
In the 3.5--20\,keV range it shows a slight decrease from 0.4 to 0.3. Towards higher energies the pulsed fraction increases significantly up to 0.9 at 50\,keV. Most notably, the Fe\,K$\alpha$ line shows far weaker pulsations than other soft components as the pulsed fraction decreases from 0.4 to ${\sim}$0.1 at the energy of the iron emission line. Such a behavior was previously reported by \citet[Fig.~7 therein]{fuerst2011} based on a \textit{XMM}-Newton observation of \gx during the pre-periastron flare.

The main focus of our phase-resolved analysis is to assess the presence of the two cyclotron lines and if present determine their pulse-phase variability. For this purpose, we fit our model as in  Eq.~\eqref{eq:baseline} to eight phase bins that are extracted from the data of the whole observation. 

\begin{figure*}
    \centering
    \resizebox{\hsize}{!}{\includegraphics{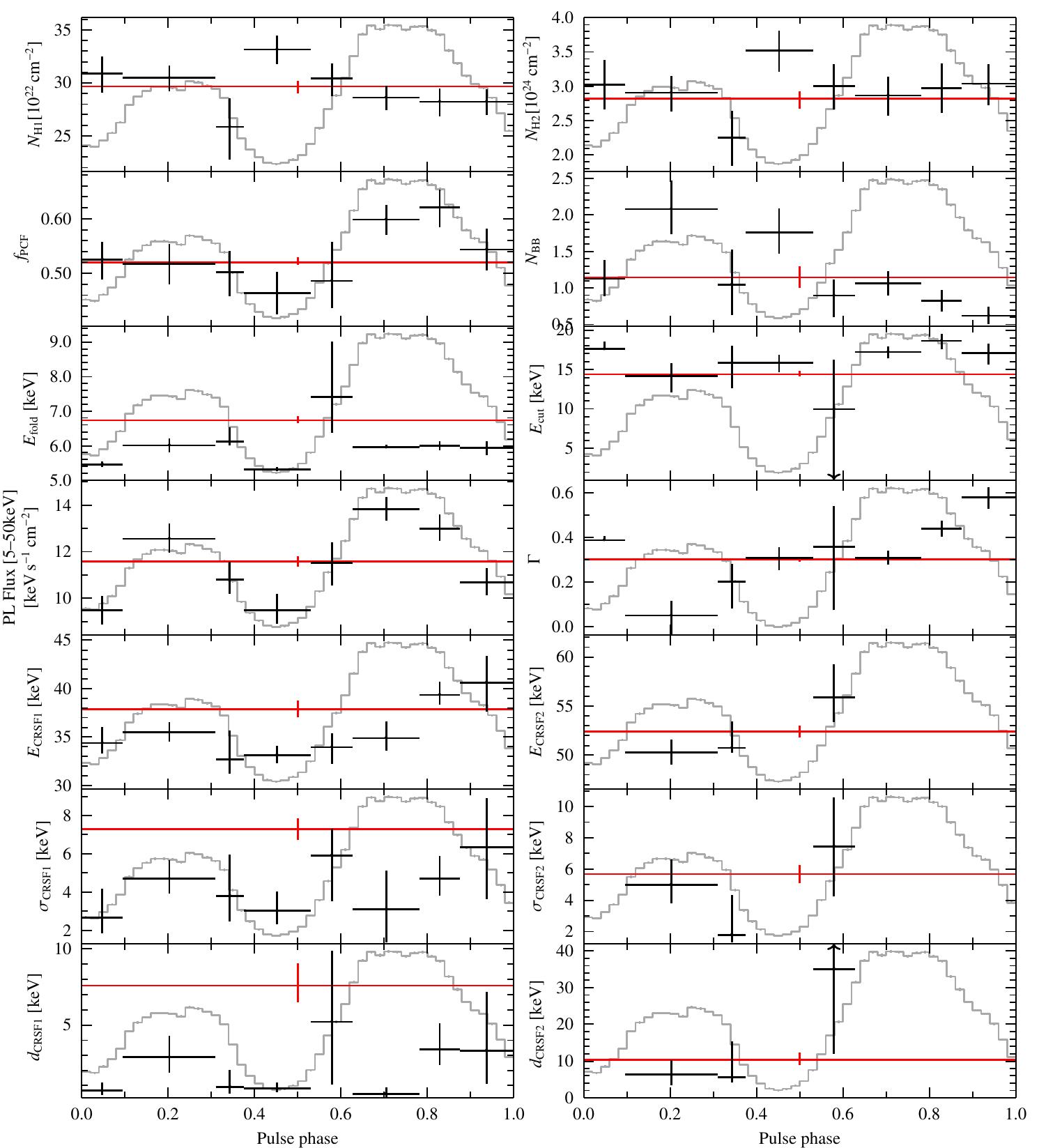}}
    \caption{Phase dependencies of all free parameters in the phase-averaged analysis. Red: Phase-averaged parameter values. Gray: Normalized pulse profile. Arrows indicate where uncertainty intervals hit their predefined range limits.}
    \label{fig:prs_phDep}
\end{figure*}

\begin{figure}
    \centering
    \resizebox{\hsize}{!}{\includegraphics[width=\linewidth]{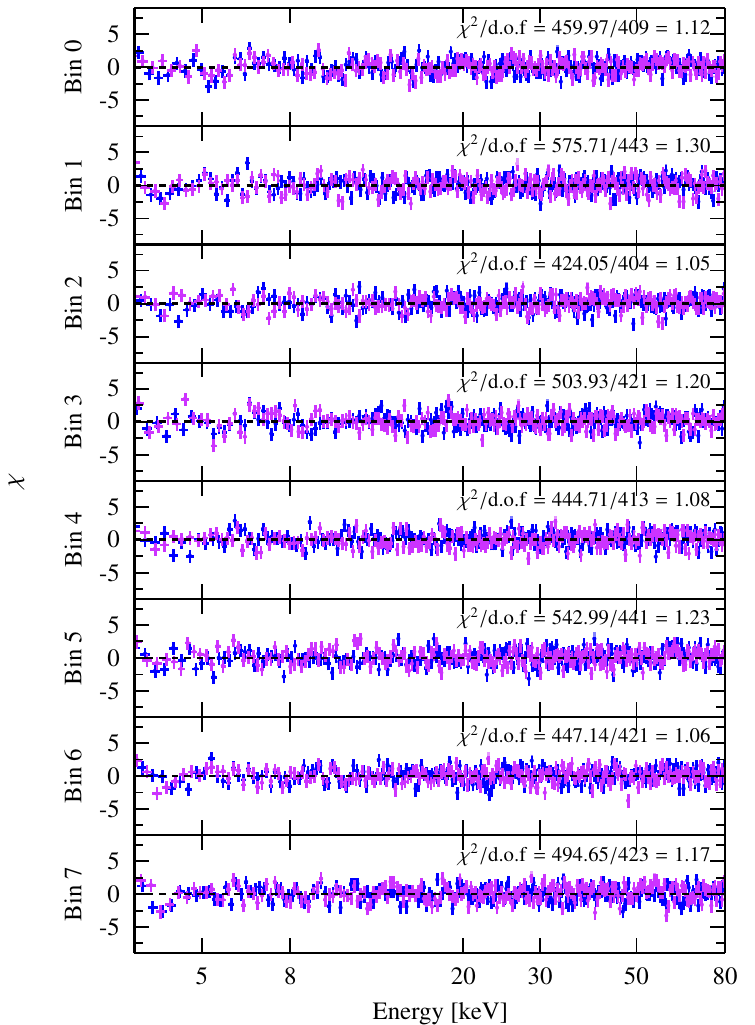}}
    \caption{$\chi$ residuals of fits to the phase-resolved spectra. Color-coding of \nustar focal plane modules as in above figures.}
    \label{fig:PRSresidualplot}
\end{figure}

We choose the phase bins by considering both the pulse profile and hardness ratios, where we aim to divide bins wherever significant changes in either quantity are present, as marked in Fig.~\ref{fig:prs_ppHR} with vertical dashed lines. The phase-resolved background spectra contain very few counts due to their respective short exposure times. As the background is not expected to vary with pulse phase, we use the phase-averaged background spectra also for the phase-resolved analysis to improve statistics. We freeze parameters which are not expected to vary with pulse-phase and showed little to no phase-dependence in a preliminary analysis. They include the detector constant, gain shift, blackbody temperature, Gaussian emission lines and the Gaussian soft emission component. We show the phase dependencies of all free parameters in Fig.~\ref{fig:prs_phDep} and the residuals of the fits in Fig.~\ref{fig:PRSresidualplot}.

We find that CRSF1 is present in all eight phase bins, as residuals of fits excluding it show deviations from the model around 35\,keV. For the higher CRSF, the picture is not as clear. At its high energy, background events and the low count rate complicate the investigation of its presence. In order to nevertheless judge the presence of CRSF2 reliably multiple criteria have to be considered:
\begin{enumerate}
    \item Do the residuals indicate a missing model component between 45--60\,keV? (see Table~\ref{tab:prsCRSFConfidence}, second column)
    \item Can the fitted energy, width and strength of CRSF2 be constrained within reason?\footnote{$E_\mathrm{CRSF2}$ between 45--60\,keV, $\sigma_\mathrm{CRSF2}$ between 0--12\,keV and $d_\mathrm{CRSF2}$ between 0--50\,keV.} If not, does a fit with $E_\mathrm{CRSF2}$ and $\sigma_\mathrm{CRSF2}$ as in the phase-averaged analysis in Sect.~\ref{sec:wholeobs} and free $d_\mathrm{CRSF2}$ provide a good description of the data? (see Table~\ref{tab:prsCRSFConfidence}, third column)
    \item Does a statistical test support the detection claim of CRSF2 with at least 3$\sigma$ confidence? (see Table~\ref{tab:prsCRSFConfidence}, fourth through seventh column)
\end{enumerate}

We note that in other publications the absence of a CRSF is often statistically substantiated by showing that the strength of the fitted CRSF is consistent with zero \citep[see, e.g.,][]{Koenig2020}. This approach is however not feasible in this case due to the low signal-to-noise ratio at energies exceeding ${\sim}$50\,keV: even if no CRSF is present, the included \texttt{gabs} component models either part of the continuum and cutoff or narrow continuum troughs of stochastic nature. In either case, the fitted strength is not consistent with zero. However, in those cases the improvement in $\chi^2$ is sufficiently small to utilize the same statistical test as used for the phase-averaged analysis in Sect.~\ref{sec:wholeobs}.

We employ the above described strategy for each phase bin and summarize our findings in Table~\ref{tab:prsCRSFConfidence}. Based on our established criteria, we detect CRSF2 with at least 3$\sigma$ significance in phase bins 1, 2 and 4. We stress that the non-detection in the other phase bins can either be due to the abscence of the spectral feature during those time or insufficient count rates to detect the spectral feature. We therefore only claim the detection of CRSF2 in phase bins 1,2 and 4, but not that CRSF2 is absent in the other phase bins.

\begin{table*}
    \caption{Summary of search for CRSF2 in the phase-resolved spectra.}
    \begin{tabular}{cccrrrrc}
    \hline \multirow{2}{*}{Phase Bin} & \multirow{2}{*}{Residuals indicate CRSF2 \tablefootmark{a)}} & \multirow{2}{*}{$E,\sigma$ constrained\tablefootmark{b)}} & \multirow{2}{*}{$\Delta \chi^2$\,\tablefootmark{c)}} & \multirow{2}{*}{$N_\mathrm{FP}$ \tablefootmark{d)}} & \multicolumn{2}{c}{Confidence} & \multirow{2}{*}{Final evaluation\tablefootmark{e)}} \\
    &&&&& [\%] & [$\sigma$] & \\\hline
    0 & N & N & 6 & 15\,013 & 84.987 & 1.4$\sigma$ & N \\
1 & N & Y & 27 & 38 & 99.962 & 3.5$\sigma$ & Y \\
2 & Y & Y & 50 & 5 & 99.995 & 4.0$\sigma$ & Y \\
3 & N & N & 3 & 50\,986 & 49.014 & 0.6$\sigma$ & N \\
4 & Y & Y & 40 & 12 & 99.988 & 3.8$\sigma$ & Y \\
5 & N & Y & 7 & 17\,212 & 82.788 & 1.3$\sigma$ & N \\
6 & N & N & 3 & 35\,919 & 64.081 & 0.9$\sigma$ & N \\
7 & N & Y & 8 & 4\,161 & 95.839 & 2.0$\sigma$ & N \\ \hline     \end{tabular} \\
    \tablefoottext{a}{Indicator whether the residuals of a fit which excludes CRSF2 indicate a missing model component at 50\,keV.} \\
    \tablefoottext{b}{Can the energy, width and strength of CRSF2 be constrained in a spectral fit?}\\
    \tablefoottext{c}{Improvement in statistic upon inclusion of CRSF2 in the spectral model.}\\
    \tablefoottext{d}{Number of false positives out of 100\,000 simulated spectra where $\Delta \chi^2$ exceeds the corresponding value upon inclusion of CRSF2.}\\
    \tablefoottext{e}{Final evaluation whether CRSF2 is detected in the corresponding phase bin.}
    \label{tab:prsCRSFConfidence}
\end{table*}

Both absorption column densities show little variation with pulse-phase and are evenly distributed around the value obtained in the phase-averaged analysis. The partial covering fraction traces the pulse profile. The blackbody component is also variable over the pulse-phase and decreases with pulse-phase apart from two outliers in phase bins 1 and 3. Apart from one outlier in phase bin 4, the folding energy is consistently below the phase-averaged value and is variable in the range 5.5--6.5\,keV. To investigate why the phase-resolved values are not evenly distributed around the phase-averaged value we remind that the continuum and CRSF1 parameters in Sect.~\ref{sec:wholeobs} changed significantly upon the inclusion of CRSF2. In a fit of the phase-averaged spectrum with only CRSF1 present, we find a folding energy of $5.812^{+0.049}_{-0.022}$\,keV; we find that the phase-resolved values for $E_\mathrm{fold}$ are distributed evenly around it. The cutoff energy is consistently above the phase-averaged value and varies with pulse-phase, for which a similar argument as for the folding energy can be made. Its lower confidence limit cannot be constrained in phase bin 4, which is most likely the result of an interplay with CRSF2, whose strength is significantly overestimated in this phase bin and also cannot be fully constrained. The powerlaw flux varies with pulse-phase as expected; the photon index $\Gamma$ also shows significant pulse-phase variability. For CRSF1 we find that the line energy is generally lower than that found in the phase-averaged analysis and increases from ${\sim} 32$\,keV to ${\sim} 41$\,keV from phases 0.4--1.0. The width of CRSF1 is lower than the value obtained in the phase-averaged analysis and is consistent with a value of ${\sim}4\,$keV. The strength is also notably weaker; the line appears to be shallow, but continuously detectable. For CRSF2, the line energy increases over the three phase bins during which it was detected. The widths scatter around the value determined in the phase-averaged analysis. The strength in phase bins 1 and 2 is consistent with the phase-averaged value, in phase bin 4 however its upper limit cannot be constrained and the parameter value itself most likely significantly overestimates its actual contribution due to correlations with the continuum parameters, in particular the cutoff and folding energies.

In summary, whilst CRSF1 is detectable over the whole pulse period, CRSF2 can only be detected with high statistical significance during a fraction of the pulse period.

\section{Discussion} \label{sec:discussion}
In the discussion of our findings we focus on the properties of the detected cyclotron lines. In Sect.~\ref{sec:disc_var}, we discuss the phase-variability of the cyclotron lines. In Sect.~\ref{sec:crsf_nature} we discuss the relation between the two cyclotron lines and their potential independence from each other. In Sect.~\ref{sec:lineformingregion} we interpret our findings from a physical point of view and propose scenarios for the formation of both lines. In Sect.~\ref{sec:crsflum} we discuss the luminosity-dependence of both CRSFs. In Sect.~\ref{sec:emissionWings} we discuss an alternative interpretation of the hard X-ray spectrum of \gx by considering emission wings of cyclotron lines.

\subsection{Phase-variability of the CRSFs}
\label{sec:disc_var}
\begin{figure}
    \centering
    \resizebox{\hsize}{!}{\includegraphics{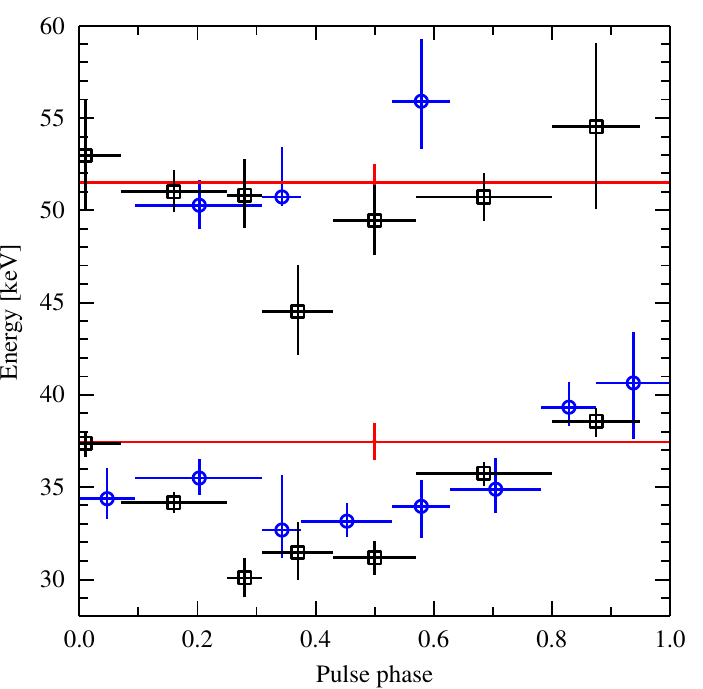}}
    \caption{Phase-dependencies of both cyclotron lines as determined in this work (blue circles) and by \citet[black squares]{Fuerst2018}. We also show our phase-averaged values in red.}
    \label{fig:prs_both_lines}
\end{figure}

In this section we aim to compare the phase-dependencies of the cyclotron line energies we observed with those obtained by \citet{Fuerst2018}. We show the cyclotron line energies as a function of pulse-phase for both this work as well as \citet{Fuerst2018} in Fig.~\ref{fig:prs_both_lines}.

For the lower line we find that our obtained values are in very good agreement with those obtained for the previous observation. For the higher line, we find that our obtained values for phase bins 1 and 2 are in good agreement whereas the value for phase bin 4 is a few keV higher in comparison. 

In order to further assess the pulse-phase-dependency of the newly discovered line it is desirable to determine its dependency at different orbital phases. \citet{Ding2021} conducted an orbital phase-resolved analysis using \textit{Insight}-HXMT and detected both cyclotron lines at different orbital phases. Furthermore, the low count rates at energies of 50\,keV and higher significantly impede both the analysis of the higher cyclotron line and its fitting using commonly used algorithms. Observations with longer exposure time and more sensitive X-ray observatories can solve this problem to some extent.

\subsection{Harmonicity of the CRSFs} \label{sec:crsf_nature}
We discuss the possibility of a harmonic relation between the two CRSFs, in order to discern their relative dependance. Other HMXBs such as 4U\,0115+63 \citep[see, e.g.,][]{Heindl1999,Ferrigno2009} and V0332+53 \citep{Tsygankov2006} are known to exhibit multiple harmonics of a CRSF in their X-ray spectra, with the harmonic lines exhibiting nearly integer multiples of the fundamental line energy. Some anharmonicity of the lines is expected from relativistic theory, as given by the resonance condition \citep[see, e.g., Eq.~1 of][]{Araya1999}
\begin{equation}
    \hbar\omega_n = \frac{\sqrt{1+2n\left(B/B_\mathrm{c}\right)\sin^2\theta}-1}{\sin^2\theta}m_\mathrm{e}c^2,
    \label{eq:cycn}
\end{equation}
where $n$ is the number of a cyclotron harmonic ($n=1$ corresponds to the fundamental, which is considered the first harmonic), $B_\mathrm{c}\approx4.413\times10^{13}\,\mathrm{G}$ is the critical field given by the Schwinger limit, $\theta$ is the viewing angle with respect to the magnetic field axis, and $m_\mathrm{e}c^2$ is the electron rest energy. For the smaller magnetic fields discussed here, with $B/B_\mathrm{c}\sim0.1$, the ratio between the second harmonic and the fundamental resonance energy is expected to be between 1.9--2.0, regardless of observer inclination. 

Here, we find a ratio between the line energies of ${\sim}1.4$, making a harmonic relation with CRSF1 as the fundamental line unlikely. We nevertheless discuss a scenario where both lines are harmonically related, as follows. 

The two observed lines could be higher harmonics of a fundamental line at ${\sim}17\,$keV, whose second and third harmonic lines would be at $34\,$keV and $51\,$keV, respectively. In this case, a ratio of ${\sim}1.4$ is expected between $\hbar\omega_3$ and $\hbar\omega_2$ (see Eq.~\ref{eq:cycn}). However, in the many analyses of the X-ray spectrum of \gx{}, such a CRSF has never been found. This is corroborated by our analysis, as the residuals displayed in Fig.~\ref{fig:phaseAveraged}b do not indicate that the description of the data between 15--20\,keV lacks a required model component. The non-detection of a CRSF at 17\,keV could however be attributed to photon spawning, which would lead to a less pronounced fundamental line. Excitation of electrons to higher Landau levels, $n\geq2$, most likely leads to radiative decay to the closest lower level, $n-1$. The energy of the emitted \enquote{spawned} photon is equal to the energy difference between the Landau levels. This leads to an excess of photons near the fundamental line energy \citep[see, e.g.,][]{Schwarm2017b,Schwarm2017a}, which could potentially fill up the otherwise observable apparent absorption feature or even lead to an excess at this energy. Given both \nustar{}'s energy resolution of 400\,eV below ${\sim}40\,$keV \citep[their Fig.~8]{Harrison2013} as well as the high luminosity of the observation, we expect that even if photon spawning plays a significant role in the filling of the fundamental line, there should be a detectable shallow absorption or weak emission feature at the fundamental line energy, as it is highly unlikely that the process of photon spawning exactly compensates the absorption feature \footnote{Photon spawning could potentially modify the shape of the fundamental line, which would displace its centroid energy, however simulations show that this effect is not pronounced \citep[see][]{Schoenherr2007, Schwarm2017b}.}. No such features were detected in this observation, or have been reported previously. 
In addition, this scenario is inconsistent with sources like 4U\,0115+63, and V0332+52, where photon spawning provides such an insignificant contribution that the clear classical ratio of harmonics is preserved to a high degree, together with a prominent fundamental line in the spectra \citep[see, e.g.,][]{Ferrigno2009}. This is made even more puzzling by the lower magnetic field of these sources which is expected to lead to higher efficiency of this process.

It is noteworthy that multiple anharmonic CRSFs have already been claimed at least once before the year 2018 for \gx: \citet{orlandini2000BeppoSAXObservationsOrbitalCycleXray} found that the \textit{BeppoSAX} spectrum of \gx requires two presumably independent CRSFs. They find time-variable line energies around 20--25\,keV and 45--55\,keV \citep[their Fig.~1]{orlandini2000BeppoSAXObservationsOrbitalCycleXray}, where the temporal variability occurs on the order of tens of days, which is most likely related to the orbital cycle of the NS and the consequent change in mass accretion rate. Their higher line can be identified as the 50\,keV line discovered by \citet{Fuerst2018}. However, \citet{Orlandini2001} later list the parameters of the higher CRSF as $E_\mathrm{CRSF}=49.5\pm1.0$\,keV and $\sigma_\mathrm{CRSF}=17.9\pm2.5\,$keV, where at least the latter parameter raises concerns on whether an actual cyclotron resonance scattering feature was detected. Such wide CRSFs have not been observed in \gx{} with \nustar{} observations and it is rather unlikely that the Doppler-broadened absorption feature would be so wide. To further illustrate this, we refer to \citet[their Eq.~1]{MeszarosNagel1985}, who give an angle-dependent formula for the width of CRSFs, which can be expressed as
\begin{equation}
    \sigma_\mathrm{CRSF}=2.55\,\mathrm{keV} \cdot \frac{E_\mathrm{CRSF}}{30\,\mathrm{keV}} \cdot \sqrt{\frac{k \,T_\mathrm{e}}{1 \, \mathrm{keV}}} \cdot \cos \theta
    \label{eq:crsfwidth}
\end{equation}
where $k$ is the Boltzmann constant, $T_\mathrm{e}$ the plasma temperature, and $\theta$ the observer's inclination to the magnetic field axis. One can see that the width stated by \citet{Orlandini2001} would require a plasma temperature of at least ${\sim}$18\,keV, which significantly exceeds the theoretically expected plasma temperature of ${\sim}$5\,keV \citep[see, e.g.,][]{Becker2007} as well as experimental findings such as ${\sim}$4--10\,keV for A\,0535+262 \citep[see][]{Ballhausen2017}. Moreover, for most sources exhibiting electron cyclotron lines the width of the features are well below ${\sim}$12\, keV \citep[see, e.g.][Table~A.5]{Staubert2019}. The preliminary analysis of a more recent \nustar observation of \gx (McCulloch et al., in prep.) shows that both CRSFs are present in the spectrum at energies consistent with values stated by \citet{Fuerst2018}.

We further note that consequent anharmonically-spaced cyclotron lines have been observed in the spectra of other accreting X-ray pulsars and have nonetheless been interpreted as harmonics. For example, a second cyclotron line was observed in the spectra of Cep~X-4, where a known fundamental line at ${\sim}30$\,keV \citep{Mihara1991} was supplemented by a harmonic at ${\sim}45$\,keV \citep[from a \textit{Suzaku} observation of the 2014 outburst, see][]{Jaisawal2015} and ${\sim}55$\,keV \citep[from a \nustar observation of the same outburst, see][]{Vybornov2017}. Similarly, in Cen~X-3, a second harmonic at ${\sim}47$\,keV was recently claimed by \citet{Yang2023} from an \textit{Insight}-HXMT observation, in addition to a previously discovered line at ${\sim}28$\,keV \citep{Audley1997, Santangelo1998}. A fundamental and a second harmonic were found at ${\sim}44$\,keV and ${\sim}73$\,keV in the spectrum of MAXI~J1409$-$619 \citep{Orlandini2012}.

\citet{Jaisawal2015} invoked a possible theoretical explanation for the non-harmonic ratio of cyclotron lines observed in \hbox{Cep~X-4}. \citet{Nishimura2005} and \citet{Schoenherr2007} showed that if the magnetic field strength varies within the line-forming region, the ratio of subsequent harmonics deviates from 1.9, and instead, the ratio increases for magnetic fields that decrease with the height of the line-forming region and vice versa \citep[see also][]{Kumar2022}. According to \citet{Schoenherr2007}, spawned photons play a major role in this process, providing more efficient filling of the fundamental line.  

Cyclotron line energies have been found to vary both with time and luminosity. For Vela~X-1, \citet{Fuerst2014} find a positive correlation of the second harmonic line whilst the fundamental line shows a negative correlation for luminosities below ${\sim} 7 \cdot 10^{36}\,\mathrm{erg}\,\mathrm{s^{-1}}$, followed by a flattening up to ${\sim} 10^{37}\,\mathrm{erg}\,\mathrm{s^{-1}}$. This implies that there is a direct proportionality between line energy ratios and luminosity, and therefore that the ratios of cyclotron line energies are also not necessarily constant over time. More recently, \citet{Diez2022} also report variability in the cyclotron line energies of both the first and second harmonic, but could not confirm an overall positive correlation of the cyclotron energies with luminosity. On the other hand, \citet{Liu2022} find a positive luminosity correlation for the fundamental line using \textit{Insight}-HXMT observations, whilst the second harmonic appears to be uncorrelated with luminosity. Based on 11 observations taken over 1.5 years, they report a decrease in the line energy ratio from ${\sim}2.0$ to ${\sim}1.7$. 

Despite the discrepancies, the aforementioned studies all agree on varying line energy ratios over time, furthermore providing a range of $\omega_2/\omega_1 {\sim}1.5$--$1.7$. Although this is slightly higher than our obtained line energy ratio of ${\sim }1.4$, they are significantly lower than the ${\sim}1.9$ expected from simple theoretical considerations. We display the ratio between the cyclotron line energies $E_\mathrm{CRSF2}/E_\mathrm{CRSF1}$ of \gx as stated in recent publications in Table~\ref{tab:ratiotab}. It is evident that our line energy ratio is the lowest to date, while the other reported values lie within the range observed in other sources. Our obtained value is 
in agreement with the two line energy ratios reported by \citet{Fuerst2018} within uncertainties. As indicated by the examples from literature above, a non-integer line energy ratio alone is no convincing argument that two cyclotron lines are not harmonically related. Below, we therefore discuss possible physical scenarios for both the harmonic and anharmonic case.

\begin{table}
    \caption{Ratio of cyclotron line energies in recent publications on \gx.}
        \label{tab:ratiotab}   
    \renewcommand{\arraystretch}{1.3}
\begin{tabular}{lrr}
         \hline Reference & $\varphi_\mathrm{orb}$\,\tablefootmark{a)} & Ratio \\ \hline 
This work & $0.97$ & $1.38\pm0.07$ \\ 
\citet{Ding2021} & $0.09$ & $1.68^{+0.12}_{-0.08}$ \\ 
\citet{Nabizadeh2019} & $0.93$ & $1.501^{+0.032}_{-0.029}$ \\ 
\citet{Fuerst2018}\,\tablefootmark{b)} & $0.67$ & $1.46\pm0.13$ \\ 
\citet{Fuerst2018}\,\tablefootmark{b)} & $0.87$ & $1.44^{+0.11}_{-0.10}$ \\ \hline     \end{tabular}\\
    \tablefoottext{a}{Orbital phase, which is calculated in the same manner as described in the caption of Fig.~\ref{fig:orbitPlot}.}\\
    \tablefoottext{b}{\citet{Fuerst2018} analyze two \nustar observations, hence the two separate table entries.}
    
\end{table}

\subsection{Line-forming regions and physical interpretations} 
\label{sec:lineformingregion}
In consideration of the results of the phase-averaged analysis in Sect.~\ref{sec:pha}, most notably the non-integer line energy ratio, and the phase-resolved analysis in Sect.~\ref{sec:phaseresolved}, which showed that CRSF2 is only detectable during a fraction of the pulse phase, we here discuss possible formation scenarios for the two CRSFs in both the harmonic and anharmonic case. We further debate multiple physical scenarios which could result in this behavior.

Since CRSFs serve as probes for the magnetic field strength of the NS, the presence of two cyclotron lines in \gx{}'s spectrum implies that if independently formed, they must form in regions with different magnetic field strengths. From \citet{Staubert2019}, Eq.~1 therein, we can estimate the field strengths to be $B_1 \approx 4.2 \times 10^{12}\,$G, and $B_2 \approx 5.8 \times 10^{12}\,$G, in the corresponding line-forming regions. In the following, we briefly discuss possible explanations for the observed CRSFs and where they are formed. First, we revisit \citet{Fuerst2018}'s interpretation that the high-energy line is formed at the NS surface whereas the lower line is formed at a height of ${\sim}$1.5\,km. \citeauthor{Fuerst2018} argue that no phase-variability is expected for CRSF2 in this case, which is corroborated by their analysis. However, we would like to note that due to the strong dependence of the cyclotron resonance profile with respect to the angle between observer inclination and the magnetic field axis features formed at the surface of the NS are still expected to show significant phase-variability \citep[see, e.g.][Fig.~5]{Schwarm2017b}. Our analysis instead shows CRSF1 present during all pulse phases and CRSF2 only present in phase bins 1, 2 and~4. However, this discrepancy can be resolved, given the following geometry: the orientation of the accretion column could be such that its bottom is obstructed for phases 0 and 5--8 (neglecting phase bin 3), resulting in the non-detection of CRSF2 there. CRSF1 however is expected to form at an altitude of 1.5\,km above the surface and can either be observed directly without obstruction by the neutron star or aided by relativistic light bending around the neutron star \citep[see, e.g.,][]{Falkner2018}. In conclusion, there indeed exists a scenario where two independently formed cyclotron lines could have formed within the same accretion column, and be in agreement with the observed pulse-phase variability. 

Further, we consider the possibility that the two lines form within different accretion columns situated at the magnetic poles of the NS. In this picture, the observed phenomenon can be understood based on arguments related to the system geometry or magnetic field structure. For pulsars, we expect that the position of the accretion columns does not coincide with the rotational axis of the NS and are separated by an offset angle $\chi$. For such inclined rotators, the magnetospheric radius decreases with increasing offset angle \citep{Bozzo2018}.
This could result in different mass-accretion rates onto the poles of the NS. In this case, the CRSFs may form at different heights within the two columns. In such a scenario, these two independently formed CRSFs can be observed as anharmonic lines with differing pulse-phase variability.

Similarly, even with symmetric mass accretion rates an explanation for the observed cyclotron line behaviour can be found: magnetic field structures can be more complex than the often assumed ideal dipole field, e.g., a distorted dipole field, which has been invoked to explain asymmetric pulse profiles as shown in Fig.~\ref{fig:prs_ppHR} \citep{Kraus1995}.
If present, such an asymmetric magnetic field would naturally lead to differing magnetic field strengths in line-forming regions within the accretion columns, which subsequently leads to the formation of CRSFs with differing centroid energies.  \citet{Jaisawal2015} suggest that a significant phase shift in CRSF parameters is expected in this case.

A distorted dipole field has already been invoked for other systems. For example, \citet{Liu2020} detected two fundamental CRSFs at 12\,keV and 16\,keV in the X-ray spectrum of 4U\,0115+63, and based on the phase-variability of the line equivalent widths, argued that the lines each form in different accretion columns, in regions with magnetic field strengths of ${\sim}1.1 \times 10^{12}$\,G and ${\sim}1.4 \times 10^{12}$\,G. The small difference in line energy makes it difficult to disentangle the higher harmonics of each fundamental line \citet[][Sect.~4.1]{Liu2020}.

The results of our analyses of \gx{} presented here can be explained by any of the aforementioned scenarios and further research is required to definitively determine whether the two cyclotron lines form within the same accretion column and if so, whether they are harmonically related.

\subsection{Cyclotron line energies in historic context}
\label{sec:crsflum}
Since the first claim of a cyclotron line in the spectrum of \gx by \citet{Mihara1995b} nearly 30 years ago, many subsequent analyses reported cyclotron lines at energies in the range 30--55\,keV. In this section we collate many CRSF energy values for \gx from literature and investigate possible explanations for the differing values in the literature.  We further aim to determine whether a possible luminosity-correlation of the CRSF line energy can account for them. Many cyclotron line sources are known to show such a correlation \citep[see, e.g.,][]{Staubert2019}. For \gx{} such a dependency was suggested by \citet{LaBarbera2005}. 

\begin{figure}
    \centering
    \resizebox{\hsize}{!}
    {\includegraphics{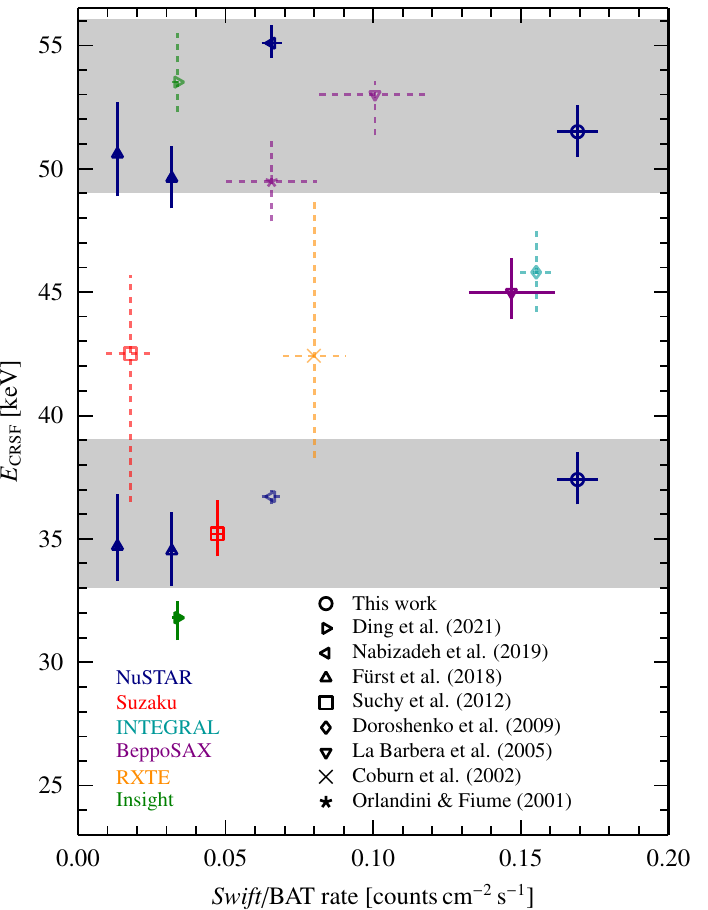}}
    \caption{Cyclotron line energy as a function of \bat count rate. For the data points from this work we use the CRSF line energies as obtained from the phase-averaged analysis in Sect.~\ref{sec:wholeobs}. Where required, the ASM count rate was converted as \bat count rate as described in Appendix~\ref{sec:AppendixA}. The uncertainties are given on a 90\% level. The semi-transparent data points with dashed error bars do not satisfy the criteria described in Sect.~\ref{sec:crsflum}, as they are either too wide or the line energy is too poorly constrained. The two gray areas represent regions where we find the cyclotron line energies to be believable. Data are from \citet{Ding2021},
    \citet{Nabizadeh2019}, \citet{Fuerst2018},
    \citet{Suchy2012}, \citet{Doroshenko2010},
    \citet{LaBarbera2005}, \citet{coburn2002MagneticFieldsAccretingRayPulsars}, and \citet{Orlandini2001}.}
    \label{fig:CRSFlum}
\end{figure}

As the luminosity and flux estimates stated in relevant publications naturally depend on the spectral model used and the energy range of the instrument, we resort to different means to estimate the source brightness. Both the \bat \citep{Krimm2013} and the \textit{RXTE} ASM \citep{levine1996FirstResultsAllSkyMonitorITAL} observatories have been monitoring \gx{} almost daily for more than 25 years \citep[for an extensive analysis of the ASM light curve  of \gx see][]{Leahy2002}. A method to allow comparison between the observed count rates from the two observatories is given in Appendix~\ref{sec:AppendixA}. We take the \bat count rate as a rough luminosity estimate and consider the variation of CRSF energy with respect to it. One caveat of this method is the strong assumption that the source spectrum remains constant over time, which is generally not justified. We note however that \citet{Staubert2014,Staubert2016} employ the same method in their analysis of the long-term decay of the cyclotron line energy in Her~X-1. We show the obtained data points in Fig. \ref{fig:CRSFlum}.

Fig.~\ref{fig:CRSFlum} shows that no simple luminosity-correlation can be inferred. Moreover, we note that CRSFs have been found both within the known ranges of 30--40\,keV and 50--55\,keV as well as between the two intervals, where we note that CRSF energies are generally less well constrained. Furthermore, it appears that observations with \nustar{} and \textit{Insight}-HXMT can constrain the CRSF line energy much better compared to observations with other instruments. Our analysis above showed correlations between CRSF and continuum parameters which raises doubts whether all data points shown in Fig.~\ref{fig:CRSFlum} correctly model an absorption component interpreted as a cyclotron line. Therefore, we only consider CRSF values that fulfill the following criteria:
\begin{enumerate}
    \item The ratio between uncertainty in line energy and line energy $\Delta E_\mathrm{CRSF}/E_\mathrm{CRSF}$ must be at most 20\%.
    \item The width of the CRSF must be reasonably small. We estimate an upper limit of the CRSF width by using Eq.~\eqref{eq:crsfwidth} for $\theta=0$, $E_\mathrm{CRSF}$ as stated in the literature and $kT_\mathrm{e}=6.50\,$keV as obtained in Sect.~\ref{sec:wholeobs} for the folding energy, which serves as a rough estimate of the plasma temperature. We discard all CRSF values whose width exceeds this value by more than $20\%$.
\end{enumerate}
The CRSF values that meet the criteria are shown in Fig.~\ref{fig:CRSFlum} as opaque data points with solid error bars, while values not meeting the criteria are shown as semi-transparent data points with dashed error bars. Our classification mostly removes CRSF values that are between the two well-known \enquote{bands}. A clear correlation with luminosity in terms of BAT count rate can however still not be inferred from our collected and filtered data points.

\subsection{Models including cyclotron line emission wings}\label{sec:emissionWings}
In the above sections, we discussed spectral models utilizing two cyclotron line components. There are however physical models that can potentially explain the observed X-ray spectrum of \gx using only one such feature. \citet{Schoenherr2007} studied cyclotron line shapes in detail and state that CRSFs often exhibit emission wings. If emission wings are present in the spectrum, the neglect to model them can lead to wavy residuals, from which additional cyclotron lines could erroneously be identified. Recently, it was shown that for low mass-accretion rates, emission wings of a cyclotron line formed by resonant redistribution in the hot NS atmosphere can significantly modify a wide range of the X-ray continuum \citep{SokolovaLapa2021, Mushtukov2021}.  This process has likely led to erroneous interpretation of a dip in the spectrum of Be X-ray binary X~Persei as a cyclotron line \citep{Coburn2001}. Similar dips were shown to be present at intermediate energies in low-luminosity observations of GX\,304$-$1 \citep{Tsygankov2019a} and A\,0535+262 \citep{Tsygankov2019b}, for which cyclotron lines are well-known at higher energies \citep[see, e.g.,][]{Malacaria2015, Ballhausen2017}. At higher mass-accretion rates, relevant for \gx, the formation of emission wings likely has less dramatic manifestation, but redistribution during Compton scattering in a warm plasma should still contribute to their formation \citep{Schoenherr2007, Schwarm2017b}. A similar nature can be attributed to the high-energy hump below the cyclotron line in the spectra of A\,0535+262 \citep{Ballhausen2017}. In addition, it was recently shown by \citet{Loudas2024} that pronounced blue wings of cyclotron lines can be formed by bulk Comptonization in the radiative shock.

\begin{figure}
    \centering
    \resizebox{\hsize}{!}{\includegraphics{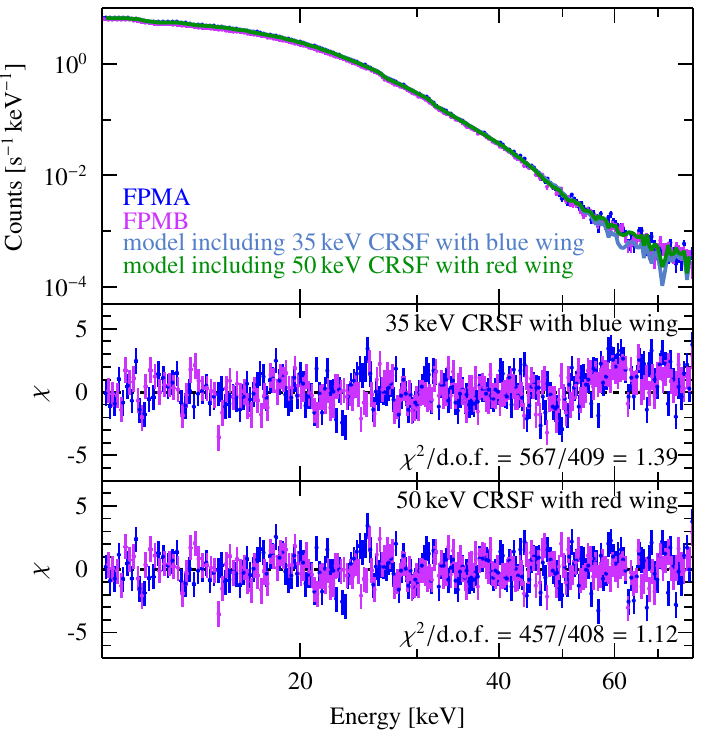}}
    \caption{Spectral fits of models including either the 35\,keV cyclotron line with blue emission wing or the 50\,keV cyclotron line with red emission wing. Top panel: 10--79\,keV \nustar data and models. Lower panels: residuals of the shown models.}
    \label{fig:emission}
\end{figure}

Here, we investigate the two possibilities that the complex structure of the continuum of \gx is created by either the 35\,keV cyclotron line with a blue emission wing situated at ${\sim}$40\,keV or the 50\,keV CRSF with a red emission wing also at ${\sim}$40\,keV. 
We therefore fit a model of the form
\begin{align}
\begin{split}
    f(E)=&\texttt{detconst}\times \texttt{PCF}(E) \times \texttt{gabs}(E) \times \\ &\Bigl(\texttt{powerlaw}(E) \times \texttt{FDcut}(E)
    + \texttt{egauss}(E) \Bigr)
\end{split}
\end{align}
to the 10--79\,keV spectrum of \gx. The parameters of the partial covering absorption component are set to those determined in Sect.~\ref{sec:wholeobs}. The spectra along with residuals are shown in Fig.~\ref{fig:emission}, the fit parameters given in Table~\ref{tab:emission_tab}.

\begin{table}
    \caption{Best-fit parameters of fits shown in Fig.~\ref{fig:emission}. Parameters without uncertainties denote fixed parameters.}
\renewcommand{\arraystretch}{1.3}
    \begin{tabular}{lrr}
        \hline Parameter & Blue Wing Fit & Red Wing Fit \\ \hline
        $K_\mathrm{PL}$\tablefootmark{a)} & $0.213\pm0.010$ & $0.179\pm0.020$\\ 
$\Gamma$ & $0.441\pm0.023$ & $0.29^{+0.07}_{-0.08}$\\ 
$E_\mathrm{fold}$ [keV] & $5.862^{+0.021}_{-0.019}$ & $7.0^{+0.7}_{-0.4}$\\ 
$E_\mathrm{cutoff}$ [keV] & $16.99\pm0.22$ & $14.2^{+2.6}_{-1.4}$\\ 
$A_\mathrm{Wing}$ [ph\,s$^{-1}$\,cm$^{-2}$] & $\left(5^{+19}_{-5}\right)\times10^{-5}$ & $\left(3.7^{+1.8}_{-1.1}\right)\times10^{-3}$\\ 
$E_\mathrm{Wing}$ [keV] & $38.23$ & $40.6\pm0.6$\\ 
$\sigma_\mathrm{Wing}$ [keV] & $0.4^{+2.4}_{-0.5}$ & $3.9^{+0.7}_{-0.6}$\\ 
$E_\mathrm{CRSF}$ [keV] & $34.6^{+1.0}_{-0.4}$ & $48.8^{+3.0}_{-2.1}$\\ 
$\sigma_\mathrm{CRSF}$ [keV] & $3.6^{+0.8}_{-0.4}$ & $12.7^{+4.1}_{-2.3}$\\ 
$d_\mathrm{CRSF}$ [keV] & $0.83^{+0.23}_{-0.12}$ & $36^{+37}_{-13}$\\ 
$C_\mathrm{FPMB}$ & $0.9989\pm0.0015$ & $0.9988\pm0.0015$\\ \hline 
$\chi^2/\mathrm{d.o.f.}$ & $567/409$ & $457/408$\\ 
$\chi^2_\mathrm{red}$ & $1.39$ & $1.12$\\ \hline 
     \end{tabular}\\
    \tablefoottext{a}{Normalization constant of powerlaw component.}\\ 
    \label{tab:emission_tab}
\end{table}

For the model of the 35\,keV CRSF, the \texttt{gabs} component naturally describes the CRSF and the \texttt{egauss} component its blue wing. The centroid energy of the blue emission wing cannot be constrained, we therefore tie it to the CRSF parameters via $E_\mathrm{Wing}=E_\mathrm{CRSF}+\sigma_\mathrm{CRSF}$. The second column of Table~\ref{tab:emission_tab} shows that the area of the blue wing is consistent with zero and its width is unexpectedly small. Along with the fact that the residuals (Fig.~\ref{fig:emission}, middle panel) still show a significant dip at 50\,keV, we conclude that this model cannot describe the X-ray continuum of \gx satisfactorily.

The model including the 50\,keV CRSF and its red wing on the other hand provides a better description of the data: it can constrain both emission wing and CRSF parameters and the residuals show no significant deviations from the data; we especially stress that no dip at 35\,keV, the energy of the prominent cyclotron line of \gx, is evident. Also in terms of the statistical evaluation of goodness of fit it provides an acceptable result. The strength of the CRSF of 36\,keV is larger than found in the analyses in Sect.~\ref{sec:pha}, which itself is however no ground for concern, as the models are built on different underlying physical ideas (multiple CRSFs versus one CRSF including emission wings). We note that the width of the CRSF of 12.7\,keV is also high in comparison and at least raises some concern whether an actual absorption feature due to cyclotron resonant scattering is being modeled \citep[see, e.g.][Table~A.5 for typical CRSF widths]{Staubert2019}. In order to follow up on this interesting and promising alternative interpretation of the observation better constraints on the high-energy emission of \gx are required.

\section{Conclusion}\label{sec:conclusion}
In summary, our analysis of the \nustar{} observation of \gx{} has shown that
\begin{enumerate}
    \item Two absorption features, most likely CRSFs, are present in both the phase-averaged and -resolved spectra. The line spacing is not harmonic and no potential fundamental line around 17\,keV has been found. Both CRSFs are present throughout the whole observation.
    \item The lower cyclotron line is present during the whole pulse period; the higher cyclotron line only during phases $\varphi \sim 0.2$--$0.6$. The pulse-phase-dependency of the lower line is in very good agreement with results by \citet{Fuerst2018}. In the phase-resolved analysis, both lines are generally more narrow and weaker than in the phase-averaged spectra.
    \item Although the ratio between the line energies of 1.38 does itself not suggest a harmonic relation between the two lines, we cannot rule out this possibility on the grounds of multiple studies of other sources, where ratios below 2 have been found. We note however, that the ratio found here is the lowest reported for \gx{} to date.
    \item Based on observations of \gx by various observatories over the last 20 years, no clear correlation of the cyclotron line energy with \bat count rate can be inferred. While there are also publications in which values for $E_\mathrm{CRSF}$ between 40--50\,keV have been found, these have to be considered in more detail as it is possible that they describe two CRSFs with one model component. Only the analysis of archival data can allow to draw conclusions about these certain values.
    \item An alternative interpretation, in which the hard X-ray spectrum is modeled with a powerlaw continuum and 50\,keV CRSF with a red emission wing provides a good description of the data and calls for further investigation of its applicability to the X-ray spectrum of \gx.
\end{enumerate}
Considering the long history of the spectral study of \gx, we note that further significant progress in disentangling cyclotron lines and continuum effects will require noticeably better constraints of the high-energy emission above ${\sim}60$\,keV at intermediate flux levels. New missions such as the probe-class mission concept High Energy X-ray Probe (\textit{HEX-P}) are expected to take on this particular challenge for accreting neutron stars in HMXBs \citep[][]{Ludlam2023}.

\begin{acknowledgements}
    We thank the anonymous referee for their very helpful comments that allowed us to improve the manuscript.
    We acknowledge support from Deutsche Forschungsgemeinschaft grants WI1860/11-2 and WI1860/19-1. 
    The material is based upon work supported by NASA under award number 80GSFC21M0002.
    The authors thank Lorenzo Ducci, Victoria Grinberg, and Robbie McCulloch for their collaboration and insight on \gx and
the XMAG collaboration, especially Kent Wood, for their helpful comments and productive discussions.
    This research has made use of ISIS functions (ISISscripts) provided by ECAP/Remeis observatory and MIT (\url{https://www.sternwarte.uni-erlangen.de/isis/}), of data obtained through the High Energy Astrophysics Science Archive Research Center Online Service, provided by the NASA/Goddard Space Flight Center, and of NASA’s Astrophysics Data System.
    This research has made use of the NuSTAR Data Analysis Software (NuSTARDAS) jointly developed by the ASI Science Data Center (ASDC, Italy) and the California Institute of Technology (USA).
    This work has made use of data from the European Space Agency (ESA) mission
    \textit{Gaia} (\url{https://www.cosmos.esa.int/gaia}), processed by the \textit{Gaia}
    Data Processing and Analysis Consortium (DPAC,
    \url{https://www.cosmos.esa.int/web/gaia/dpac/consortium}).
    Funding for the DPAC
    has been provided by national institutions, in particular the institutions
    participating in the \textit{Gaia} Multilateral Agreement.
\end{acknowledgements}

\bibliographystyle{aa}
\bibliography{references.bib}

\begin{thebibliography}{108}
\expandafter\ifx\csname natexlab\endcsname\relax\def\natexlab#1{#1}\fi

\bibitem[{{Alonso-Hern{\'a}ndez} {et~al.}(2022){Alonso-Hern{\'a}ndez},
  {F{\"u}rst}, {Kretschmar}, {Caballero}, \& {Joyce}}]{AlonsoHernandez2022}
{Alonso-Hern{\'a}ndez}, J., {F{\"u}rst}, F., {Kretschmar}, P., {Caballero}, I.,
  \& {Joyce}, A.~M. 2022, \aap, 662, A62

\bibitem[{{Araya} \& {Harding}(1999)}]{Araya1999}
{Araya}, R.~A. \& {Harding}, A.~K. 1999, ApJ, 517, 334

\bibitem[{{Audley}(1997)}]{Audley1997}
{Audley}, M.~D. 1997, PhD thesis, University of Maryland, College Park

\bibitem[{{Bailer-Jones} {et~al.}(2021){Bailer-Jones}, Rybizki, Fouesneau,
  Demleitner, \&
  Andrae}]{bailer-jones2021EstimatingDistancesParallaxesGeometricPhotogeometric}
{Bailer-Jones}, C. A.~L., Rybizki, J., Fouesneau, M., Demleitner, M., \&
  Andrae, R. 2021, AJ, 161, 147

\bibitem[{{Ballhausen}(2021)}]{Ballhausen2021}
{Ballhausen}, R. 2021, PhD thesis, {Friedrich-Alexander-Universit\"at
  Erlangen-N\"urnberg}

\bibitem[{{Ballhausen} {et~al.}(2020){Ballhausen}, {Lorenz}, {F{\"u}rst},
  {Pottschmidt}, {Corrales}, {Tomsick}, {Bissinger n{\'e} K{\"u}hnel},
  {Kretschmar}, {Kallman}, {Grinberg}, {Hell}, {Psadaraki}, {Rogantini}, \&
  {Wilms}}]{Ballhausen2020}
{Ballhausen}, R., {Lorenz}, M., {F{\"u}rst}, F., {et~al.} 2020, \aap, 641, A65

\bibitem[{{Ballhausen} {et~al.}(2017){Ballhausen}, {Pottschmidt}, {F{\"u}rst},
  {Wilms}, {Tomsick}, {Schwarm}, {Stern}, {Kretschmar}, {Caballero},
  {Harrison}, {Boggs}, {Christensen}, {Craig}, {Hailey}, \&
  {Zhang}}]{Ballhausen2017}
{Ballhausen}, R., {Pottschmidt}, K., {F{\"u}rst}, F., {et~al.} 2017, \aap, 608,
  A105

\bibitem[{{Bearden}(1967)}]{Bearden1967}
{Bearden}, J.~A. 1967, Rev. Mod. Phys., 39, 78

\bibitem[{{Becker} {et~al.}(2012){Becker}, {Klochkov}, {Sch{\"o}nherr},
  {Nishimura}, {Ferrigno}, {Caballero}, {Kretschmar}, {Wolff}, {Wilms}, \&
  {Staubert}}]{Becker2012}
{Becker}, P.~A., {Klochkov}, D., {Sch{\"o}nherr}, G., {et~al.} 2012, \aap, 544,
  A123

\bibitem[{{Becker} \& {Wolff}(2007)}]{Becker2007}
{Becker}, P.~A. \& {Wolff}, M.~T. 2007, \apj, 654, 435

\bibitem[{{Bozzo} {et~al.}(2018){Bozzo}, {Ascenzi}, {Ducci}, {Papitto},
  {Burderi}, \& {Stella}}]{Bozzo2018}
{Bozzo}, E., {Ascenzi}, S., {Ducci}, L., {et~al.} 2018, \aap, 617, A126

\bibitem[{{Bradt} {et~al.}(1977){Bradt}, {Apparao}, {Clark}, {Dower}, {Doxsey},
  {Hearn}, {Jernigan}, {Joss}, {Mayer}, {McClintock}, \& {Walter}}]{Bradt1977}
{Bradt}, H.~V., {Apparao}, K.~M.~V., {Clark}, G.~W., {et~al.} 1977, Nat, 269,
  21

\bibitem[{{Coburn} {et~al.}(2001){Coburn}, {Heindl}, {Gruber}, {Rothschild},
  {Staubert}, {Wilms}, \& {Kreykenbohm}}]{Coburn2001}
{Coburn}, W., {Heindl}, W.~A., {Gruber}, D.~E., {et~al.} 2001, \apj, 552, 738

\bibitem[{Coburn {et~al.}(2002)Coburn, Heindl, Rothschild, Gruber, Kreykenbohm,
  Wilms, Kretschmar, \& Staubert}]{coburn2002MagneticFieldsAccretingRayPulsars}
Coburn, W., Heindl, W.~A., Rothschild, R.~E., {et~al.} 2002, ApJ, 580, 394

\bibitem[{{Diez} {et~al.}(2023){Diez}, {Grinberg}, {F{\"u}rst}, {El Mellah},
  {Zhou}, {Santangelo}, {Mart{\'\i}nez-N{\'u}{\~n}ez}, {Amato}, {Hell}, \&
  {Kretschmar}}]{Diez2023}
{Diez}, C.~M., {Grinberg}, V., {F{\"u}rst}, F., {et~al.} 2023, \aap, 674, A147

\bibitem[{{Diez} {et~al.}(2022){Diez}, {Grinberg}, {F{\"u}rst},
  {Sokolova-Lapa}, {Santangelo}, {Wilms}, {Pottschmidt},
  {Mart{\'\i}nez-N{\'u}{\~n}ez}, {Malacaria}, \& {Kretschmar}}]{Diez2022}
{Diez}, C.~M., {Grinberg}, V., {F{\"u}rst}, F., {et~al.} 2022, \aap, 660, A19

\bibitem[{{Ding} {et~al.}(2021){Ding}, {Wang}, {Epili}, {Liu}, {Ge}, {Lu},
  {Qu}, {Song}, {Zhang}, \& {Zhang}}]{Ding2021}
{Ding}, Y.~Z., {Wang}, W., {Epili}, P.~R., {et~al.} 2021, \mnras, 506, 2712

\bibitem[{{Doroshenko} {et~al.}(2010){Doroshenko}, {Santangelo}, {Suleimanov},
  {Kreykenbohm}, {Staubert}, {Ferrigno}, \& {Klochkov}}]{Doroshenko2010}
{Doroshenko}, V., {Santangelo}, A., {Suleimanov}, V., {et~al.} 2010, \aap, 515,
  A10

\bibitem[{{Evangelista} {et~al.}(2010){Evangelista}, {Feroci}, {Costa}, {Del
  Monte}, {Donnarumma}, {Lapshov}, {Lazzarotto}, {Pacciani}, {Rapisarda},
  {Soffitta}, {Argan}, {Barbiellini}, {Boffelli}, {Bulgarelli}, {Caraveo},
  {Cattaneo}, {Chen}, {D'Ammando}, {Di Cocco}, {Fuschino}, {Galli}, {Gianotti},
  {Giuliani}, {Labanti}, {Lipari}, {Longo}, {Marisaldi}, {Mereghetti},
  {Moretti}, {Morselli}, {Pellizzoni}, {Perotti}, {Piano}, {Picozza}, {Pilia},
  {Prest}, {Pucella}, {Rappoldi}, {Sabatini}, {Striani}, {Tavani}, {Trifoglio},
  {Trois}, {Vallazza}, {Vercellone}, {Vittorini}, {Zambra}, {Antonelli},
  {Cutini}, {Pittori}, {Preger}, {Santolamazza}, {Verrecchia}, {Giommi}, \&
  {Salotti}}]{Evangelista2010}
{Evangelista}, Y., {Feroci}, M., {Costa}, E., {et~al.} 2010, \apj, 708, 1663

\bibitem[{{Falkner}(2018)}]{Falkner2018}
{Falkner}, S. 2018, PhD thesis, {Friedrich-Alexander-Universit\"at
  Erlangen-N\"urnberg}

\bibitem[{{Ferrigno} {et~al.}(2009){Ferrigno}, {Becker}, {Segreto}, {Mineo}, \&
  {Santangelo}}]{Ferrigno2009}
{Ferrigno}, C., {Becker}, P.~A., {Segreto}, A., {Mineo}, T., \& {Santangelo},
  A. 2009, \aap, 498, 825

\bibitem[{{F{\"u}rst} {et~al.}(2018){F{\"u}rst}, {Falkner}, {Marcu-Cheatham},
  {Grefenstette}, {Tomsick}, {Pottschmidt}, {Walton}, {Natalucci}, \&
  {Kretschmar}}]{Fuerst2018}
{F{\"u}rst}, F., {Falkner}, S., {Marcu-Cheatham}, D., {et~al.} 2018, \aap, 620,
  A153

\bibitem[{{F{\"u}rst} {et~al.}(2014){F{\"u}rst}, {Pottschmidt}, {Wilms},
  {Tomsick}, {Bachetti}, {Boggs}, {Christensen}, {Craig}, {Grefenstette},
  {Hailey}, {Harrison}, {Madsen}, {Miller}, {Stern}, {Walton}, \&
  {Zhang}}]{Fuerst2014}
{F{\"u}rst}, F., {Pottschmidt}, K., {Wilms}, J., {et~al.} 2014, \apj, 780, 133

\bibitem[{F\"urst {et~al.}(2011)F\"urst, {Suchy, S.}, {Kreykenbohm, I.},
  {Barrag\'an, L.}, {Wilms, J.}, {Pottschmidt, K.}, {Caballero, I.},
  {Kretschmar, P.}, {Ferrigno, C.}, \& {Rothschild, R. E.}}]{fuerst2011}
F\"urst, F., {Suchy, S.}, {Kreykenbohm, I.}, {et~al.} 2011, A\&A, 535, A9

\bibitem[{{Gaia Collaboration}(2016)}]{gaiacollaboration2016GaiaMission}
{Gaia Collaboration}. 2016, A\&A, 595, A1

\bibitem[{{Gaia
  Collaboration}(2023)}]{gaiacollaboration2022GaiaDataReleaseSummaryContent}
{Gaia Collaboration}. 2023, \aap, 674, A1

\bibitem[{{Grefenstette} {et~al.}(2022){Grefenstette}, {Brightman}, {Earnshaw},
  {Forster}, {Madsen}, \& {Miyasaka}}]{Grefenstette2022}
{Grefenstette}, B., {Brightman}, M., {Earnshaw}, H.~P., {et~al.} 2022,
  arXiv:2206.04058, technical note

\bibitem[{{Harrison} {et~al.}(2013){Harrison}, {Craig}, {Christensen},
  {Hailey}, {Zhang}, {Boggs}, {Stern}, {Cook}, {Forster}, {Giommi},
  {Grefenstette}, {Kim}, {Kitaguchi}, {Koglin}, {Madsen}, {Mao}, {Miyasaka},
  {Mori}, {Perri}, {Pivovaroff}, {Puccetti}, {Rana}, {Westergaard}, {Willis},
  {Zoglauer}, {An}, {Bachetti}, {Barri{\`e}re}, {Bellm}, {Bhalerao},
  {Brejnholt}, {Fuerst}, {Liebe}, {Markwardt}, {Nynka}, {Vogel}, {Walton},
  {Wik}, {Alexander}, {Cominsky}, {Hornschemeier}, {Hornstrup}, {Kaspi},
  {Madejski}, {Matt}, {Molendi}, {Smith}, {Tomsick}, {Ajello}, {Ballantyne},
  {Balokovi{\'c}}, {Barret}, {Bauer}, {Blandford}, {Brandt}, {Brenneman},
  {Chiang}, {Chakrabarty}, {Chenevez}, {Comastri}, {Dufour}, {Elvis}, {Fabian},
  {Farrah}, {Fryer}, {Gotthelf}, {Grindlay}, {Helfand}, {Krivonos}, {Meier},
  {Miller}, {Natalucci}, {Ogle}, {Ofek}, {Ptak}, {Reynolds}, {Rigby},
  {Tagliaferri}, {Thorsett}, {Treister}, \& {Urry}}]{Harrison2013}
{Harrison}, F.~A., {Craig}, W.~W., {Christensen}, F.~E., {et~al.} 2013, \apj,
  770, 103

\bibitem[{{Heindl} {et~al.}(1999){Heindl}, {Coburn}, {Gruber}, {Pelling},
  {Rothschild}, {Wilms}, {Pottschmidt}, \& {Staubert}}]{Heindl1999}
{Heindl}, W.~A., {Coburn}, W., {Gruber}, D.~E., {et~al.} 1999, \apjl, 521, L49

\bibitem[{{Houck} \& {Denicola}(2000)}]{Houck2000}
{Houck}, J.~C. \& {Denicola}, L.~A. 2000, in Astronomical Society of the
  Pacific Conference Series, Vol. 216, Astronomical Data Analysis Software and
  Systems IX, ed. N.~{Manset}, C.~{Veillet}, \& D.~{Crabtree}, 591

\bibitem[{{Islam} \& {Paul}(2014)}]{Islam2014}
{Islam}, N. \& {Paul}, B. 2014, \mnras, 441, 2539

\bibitem[{{Iyer} {et~al.}(2015){Iyer}, {Mukherjee}, {Dewangan}, {Bhattacharya},
  \& {Seetha}}]{Iyer2015}
{Iyer}, N., {Mukherjee}, D., {Dewangan}, G.~C., {Bhattacharya}, D., \&
  {Seetha}, S. 2015, \mnras, 454, 741

\bibitem[{{Jaisawal} \& {Naik}(2015)}]{Jaisawal2015}
{Jaisawal}, G.~K. \& {Naik}, S. 2015, \mnras, 453, L21

\bibitem[{{Kaastra} \& {Bleeker}(2016)}]{Kaastra2016}
{Kaastra}, J.~S. \& {Bleeker}, J.~A.~M. 2016, \aap, 587, A151

\bibitem[{{Kaper} {et~al.}(2006){Kaper}, {van der Meer}, \&
  {Najarro}}]{Kaper2006}
{Kaper}, L., {van der Meer}, A., \& {Najarro}, F. 2006, \aap, 457, 595

\bibitem[{{Kendziorra} {et~al.}(1992){Kendziorra}, {Mony}, {Kretschmar},
  {Maisack}, {Staubert}, {D{\"o}bereiner}, {Englauser}, {Pietsch}, {Reppin},
  {Tr{\"u}mper}, {Efremov}, {Kaniovsky}, \& {Sunyaev}}]{Kendziorra1992}
{Kendziorra}, E., {Mony}, B., {Kretschmar}, P., {et~al.} 1992, in NASA
  Conference Publication, Vol. 3137, The Compton Observatory Science Workshop,
  ed. C.~R. {Shrader}, N.~{Gehrels}, \& B.~{Dennis} (GSFC), 217

\bibitem[{Koh {et~al.}(1997)Koh, Bildsten, Chakrabarty, Nelson, Prince,
  Vaughan, Finger, Wilson, \& Rubin}]{koh1997RapidSpinEpisodesWindFed}
Koh, D.~T., Bildsten, L., Chakrabarty, D., {et~al.} 1997, ApJ, 479, 933

\bibitem[{{K{\"o}nig} {et~al.}(2020){K{\"o}nig}, {F{\"u}rst}, {Kretschmar},
  {Ballhausen}, {Sokolova-Lapa}, {Dauser}, {S{\'a}nchez-Fern{\'a}ndez},
  {Hemphill}, {Wolff}, {Pottschmidt}, \& {Wilms}}]{Koenig2020}
{K{\"o}nig}, O., {F{\"u}rst}, F., {Kretschmar}, P., {et~al.} 2020, \aap, 643,
  A128

\bibitem[{{Kraus} {et~al.}(1995){Kraus}, {Nollert}, {Ruder}, \&
  {Riffert}}]{Kraus1995}
{Kraus}, U., {Nollert}, H.~P., {Ruder}, H., \& {Riffert}, H. 1995, \apj, 450,
  763

\bibitem[{{Krause} \& {Oliver}(1979)}]{Krause1979}
{Krause}, M.~O. \& {Oliver}, J.~H. 1979, J. Phys. Chem. Ref. Data, 8, 329

\bibitem[{{Kretschmar} {et~al.}(2021){Kretschmar}, {El Mellah},
  {Mart{\'\i}nez-N{\'u}{\~n}ez}, {F{\"u}rst}, {Grinberg}, {Sander}, {van den
  Eijnden}, {Degenaar}, {Ma{\'\i}z Apell{\'a}niz}, {Jim{\'e}nez Esteban},
  {Ramos-Lerate}, \& {Utrilla}}]{Kretschmar2021}
{Kretschmar}, P., {El Mellah}, I., {Mart{\'\i}nez-N{\'u}{\~n}ez}, S., {et~al.}
  2021, \aap, 652, A95

\bibitem[{{Kretschmar} {et~al.}(1997){Kretschmar}, {Pan}, {Kendziorra},
  {Maisack}, {Staubert}, {Skinner}, {Pietsch}, {Truemper}, {Efremov}, \&
  {Sunyaev}}]{Kretschmar1997}
{Kretschmar}, P., {Pan}, H.~C., {Kendziorra}, E., {et~al.} 1997, \aap, 325, 623

\bibitem[{{Kreykenbohm} {et~al.}(2002){Kreykenbohm}, {Coburn}, {Wilms},
  {Kretschmar}, {Staubert}, {Heindl}, \& {Rothschild}}]{Kreykenbohm2002}
{Kreykenbohm}, I., {Coburn}, W., {Wilms}, J., {et~al.} 2002, \aap, 395, 129

\bibitem[{{Kreykenbohm} {et~al.}(1999){Kreykenbohm}, {Kretschmar}, {Wilms},
  {Staubert}, {Kendziorra}, {Gruber}, {Heindl}, \&
  {Rothschild}}]{Kreykenbohm1999}
{Kreykenbohm}, I., {Kretschmar}, P., {Wilms}, J., {et~al.} 1999, \aap, 341, 141

\bibitem[{{Kreykenbohm} {et~al.}(2004){Kreykenbohm}, {Wilms}, {Coburn},
  {Kuster}, {Rothschild}, {Heindl}, {Kretschmar}, \&
  {Staubert}}]{Kreykenbohm2004}
{Kreykenbohm}, I., {Wilms}, J., {Coburn}, W., {et~al.} 2004, \aap, 427, 975

\bibitem[{{Krimm} {et~al.}(2013){Krimm}, {Holland}, {Corbet}, {Pearlman},
  {Romano}, {Kennea}, {Bloom}, {Barthelmy}, {Baumgartner}, {Cummings},
  {Gehrels}, {Lien}, {Markwardt}, {Palmer}, {Sakamoto}, {Stamatikos}, \&
  {Ukwatta}}]{Krimm2013}
{Krimm}, H.~A., {Holland}, S.~T., {Corbet}, R.~H.~D., {et~al.} 2013, \apjs,
  209, 14

\bibitem[{{Kumar} {et~al.}(2022){Kumar}, {Bala}, \& {Bhattacharya}}]{Kumar2022}
{Kumar}, S., {Bala}, S., \& {Bhattacharya}, D. 2022, \mnras, 515, 914

\bibitem[{{La Barbera} {et~al.}(2005){La Barbera}, {Segreto}, {Santangelo},
  {Kreykenbohm}, \& {Orlandini}}]{LaBarbera2005}
{La Barbera}, A., {Segreto}, A., {Santangelo}, A., {Kreykenbohm}, I., \&
  {Orlandini}, M. 2005, \aap, 438, 617

\bibitem[{{Leahy}(2002)}]{Leahy2002}
{Leahy}, D.~A. 2002, \aap, 391, 219

\bibitem[{{Leahy} {et~al.}(1983){Leahy}, {Elsner}, \& {Weisskopf}}]{Leahy1983}
{Leahy}, D.~A., {Elsner}, R.~F., \& {Weisskopf}, M.~C. 1983, \apj, 272, 256

\bibitem[{{Leahy} \& {Kostka}(2008)}]{Leahy2008}
{Leahy}, D.~A. \& {Kostka}, M. 2008, \mnras, 384, 747

\bibitem[{{Levine} {et~al.}(1996){Levine}, {Bradt}, {Cui}, {Jernigan},
  {Morgan}, {Remillard}, {Shirey}, \&
  {Smith}}]{levine1996FirstResultsAllSkyMonitorITAL}
{Levine}, A.~M., {Bradt}, H., {Cui}, W., {et~al.} 1996, \apjl, 469, L33

\bibitem[{{Liu} {et~al.}(2020){Liu}, {Tao}, {Zhang}, {Li}, {Ge}, {Qu}, {Song},
  {Ji}, {Zhang}, {Santangelo}, \& {Wang}}]{Liu2020}
{Liu}, B.-S., {Tao}, L., {Zhang}, S.-N., {et~al.} 2020, \apj, 900, 41

\bibitem[{{Liu} {et~al.}(2022){Liu}, {Wang}, {Chen}, {Ding}, {Lu}, {Song},
  {Qu}, {Zhang}, \& {Zhang}}]{Liu2022}
{Liu}, Q., {Wang}, W., {Chen}, X., {et~al.} 2022, \mnras, 514, 2805

\bibitem[{{Loudas} {et~al.}(2024){Loudas}, {Kylafis}, \&
  {Tr{\"u}mper}}]{Loudas2024}
{Loudas}, N., {Kylafis}, N.~D., \& {Tr{\"u}mper}, J.~E. 2024, \aap, submitted,
  pre-print arXiv:2402.07983

\bibitem[{{Ludlam} {et~al.}(2023){Ludlam}, {Malacaria}, {Sokolova-Lapa},
  {Fuerst}, {Pradhan}, {Shaw}, {Pottschmidt}, {Pike}, {Vasilopoulos}, {Wilms},
  {Garc{\'\i}a}, {Madsen}, {Stern}, {Maitra}, {Del Santo}, {Walton},
  {Brumback}, \& {van den Eijnden}}]{Ludlam2023}
{Ludlam}, R.~M., {Malacaria}, C., {Sokolova-Lapa}, E., {et~al.} 2023, Front.
  Astron. Space Sci, 10, 1292500

\bibitem[{{Madsen} {et~al.}(2022){Madsen}, {Forster}, {Grefenstette},
  {Harrison}, \& {Miyasaka}}]{Madsen2022}
{Madsen}, K.~K., {Forster}, K., {Grefenstette}, B., {Harrison}, F.~A., \&
  {Miyasaka}, H. 2022, JATIS, 8, 034003

\bibitem[{{Maitra}(2017)}]{Maitra2017}
{Maitra}, C. 2017, J. Astrophys. Astr., 38, 50

\bibitem[{{Malacaria} {et~al.}(2023){Malacaria}, {Ducci}, {Falanga},
  {Altamirano}, {Bozzo}, {Guillot}, {Jaisawal}, {Kretschmar}, {Ng}, {Pradhan},
  {Rothschild}, {Sanna}, {Thalhammer}, \& {Wilms}}]{Malacaria2023}
{Malacaria}, C., {Ducci}, L., {Falanga}, M., {et~al.} 2023, \aap, 669, A38

\bibitem[{{Malacaria} {et~al.}(2015){Malacaria}, {Klochkov}, {Santangelo}, \&
  {Staubert}}]{Malacaria2015}
{Malacaria}, C., {Klochkov}, D., {Santangelo}, A., \& {Staubert}, R. 2015,
  \aap, 581, A121

\bibitem[{{Manikantan} {et~al.}(2024){Manikantan}, {Kumar}, {Paul}, \&
  {Rana}}]{Manikantan2023}
{Manikantan}, H., {Kumar}, M., {Paul}, B., \& {Rana}, V. 2024, \mnras, 527, 640

\bibitem[{{McClintock} {et~al.}(1971){McClintock}, {Ricker}, \&
  {Lewin}}]{McClintock1971}
{McClintock}, J.~E., {Ricker}, G.~R., \& {Lewin}, W. H.~G. 1971, \apjl, 166,
  L73

\bibitem[{{Meszaros} \& {Nagel}(1985)}]{MeszarosNagel1985}
{Meszaros}, P. \& {Nagel}, W. 1985, \apj, 299, 138

\bibitem[{{Mihara}(1995)}]{Mihara1995b}
{Mihara}, T. 1995, PhD thesis, University of Tokyo

\bibitem[{{Mihara} {et~al.}(1991){Mihara}, {Makishima}, {Kamijo}, {Ohashi},
  {Nagase}, {Tanaka}, \& {Koyama}}]{Mihara1991}
{Mihara}, T., {Makishima}, K., {Kamijo}, S., {et~al.} 1991, \apjl, 379, L61

\bibitem[{{M{\"u}ller} {et~al.}(2013){M{\"u}ller}, {Ferrigno}, {K{\"u}hnel},
  {Sch{\"o}nherr}, {Becker}, {Wolff}, {Hertel}, {Schwarm}, {Grinberg}, {Obst},
  {Caballero}, {Pottschmidt}, {F{\"u}rst}, {Kreykenbohm}, {Rothschild},
  {Hemphill}, {N{\'u}{\~n}ez}, {Torrej{\'o}n}, {Klochkov}, {Staubert}, \&
  {Wilms}}]{Mueller2013}
{M{\"u}ller}, S., {Ferrigno}, C., {K{\"u}hnel}, M., {et~al.} 2013, \aap, 551,
  A6

\bibitem[{{Mushtukov} {et~al.}(2021){Mushtukov}, {Suleimanov}, {Tsygankov}, \&
  {Portegies Zwart}}]{Mushtukov2021}
{Mushtukov}, A.~A., {Suleimanov}, V.~F., {Tsygankov}, S.~S., \& {Portegies
  Zwart}, S. 2021, \mnras, 503, 5193

\bibitem[{{Mushtukov} {et~al.}(2015){Mushtukov}, {Tsygankov}, {Serber},
  {Suleimanov}, \& {Poutanen}}]{Mushtukov2015}
{Mushtukov}, A.~A., {Tsygankov}, S.~S., {Serber}, A.~V., {Suleimanov}, V.~F.,
  \& {Poutanen}, J. 2015, \mnras, 454, 2714

\bibitem[{{Nabizadeh} {et~al.}(2019){Nabizadeh}, {M{\"o}nkk{\"o}nen},
  {Tsygankov}, {Doroshenko}, {Molkov}, \& {Poutanen}}]{Nabizadeh2019}
{Nabizadeh}, A., {M{\"o}nkk{\"o}nen}, J., {Tsygankov}, S.~S., {et~al.} 2019,
  \aap, 629, A101

\bibitem[{{Nakajima} {et~al.}(2010){Nakajima}, {Mihara}, \&
  {Makishima}}]{Nakajima2010}
{Nakajima}, M., {Mihara}, T., \& {Makishima}, K. 2010, \apj, 710, 1755

\bibitem[{{Nishimura}(2005)}]{Nishimura2005}
{Nishimura}, O. 2005, \pasj, 57, 769

\bibitem[{{Nishimura}(2014)}]{Nishimura2014}
{Nishimura}, O. 2014, \apj, 781, 30

\bibitem[{{Orlandini} {et~al.}(1998){Orlandini}, {dal Fiume}, {Frontera},
  {Cusumano}, {del Sordo}, {Giarrusso}, {Piraino}, {Segreto}, {Guainazzi}, \&
  {Piro}}]{Orlandini1998}
{Orlandini}, M., {dal Fiume}, D., {Frontera}, F., {et~al.} 1998, \aap, 332, 121

\bibitem[{Orlandini {et~al.}(2000)Orlandini, Dal~Fiume, Frontera, Oosterbroek,
  Parmar, Santangelo, \&
  Segreto}]{orlandini2000BeppoSAXObservationsOrbitalCycleXray}
Orlandini, M., Dal~Fiume, D., Frontera, F., {et~al.} 2000, Adv. Space Res., 25,
  417

\bibitem[{{Orlandini} \& {Fiume}(2001)}]{Orlandini2001}
{Orlandini}, M. \& {Fiume}, D.~D. 2001, in AIP Conference Series, Vol. 599,
  X-ray Astronomy: Stellar Endpoints, AGN, and the Diffuse X-ray Background,
  ed. N.~E. {White}, G.~{Malaguti}, \& G.~G.~C. {Palumbo} (AIP Publishing),
  283--294

\bibitem[{{Orlandini} {et~al.}(2012){Orlandini}, {Frontera}, {Masetti},
  {Sguera}, \& {Sidoli}}]{Orlandini2012}
{Orlandini}, M., {Frontera}, F., {Masetti}, N., {Sguera}, V., \& {Sidoli}, L.
  2012, \apj, 748, 86

\bibitem[{{Poutanen} {et~al.}(2013){Poutanen}, {Mushtukov}, {Suleimanov},
  {Tsygankov}, {Nagirner}, {Doroshenko}, \& {Lutovinov}}]{Poutanen2013}
{Poutanen}, J., {Mushtukov}, A.~A., {Suleimanov}, V.~F., {et~al.} 2013, \apj,
  777, 115

\bibitem[{Pradhan {et~al.}(2021)Pradhan, Paul, Bozzo, Maitra, \&
  Paul}]{Pradhan21}
Pradhan, P., Paul, B., Bozzo, E., Maitra, C., \& Paul, B.~C. 2021, MNRAS, 502,
  1163

\bibitem[{{Pravdo} {et~al.}(1995){Pravdo}, {Day}, {Angelini}, {Harmon},
  {Yoshida}, \& {Saraswat}}]{Pravdo1995}
{Pravdo}, S.~H., {Day}, C. S.~R., {Angelini}, L., {et~al.} 1995, \apj, 454, 872

\bibitem[{{Pravdo} \& {Ghosh}(2001)}]{Pravdo2001}
{Pravdo}, S.~H. \& {Ghosh}, P. 2001, \apj, 554, 383

\bibitem[{{Radhakrishnan} \& {Cooke}(1969)}]{Radhakrishan1969}
{Radhakrishnan}, V. \& {Cooke}, D.~J. 1969, \aplett, 3, 225

\bibitem[{{Roy} {et~al.}(2024){Roy}, {Manikantan}, \& {Paul}}]{Roy2023}
{Roy}, K., {Manikantan}, H., \& {Paul}, B. 2024, \mnras, 527, 2652

\bibitem[{{Santangelo} {et~al.}(1998){Santangelo}, {del Sordo}, {Segreto}, {dal
  Fiume}, {Orlandini}, \& {Piraino}}]{Santangelo1998}
{Santangelo}, A., {del Sordo}, S., {Segreto}, A., {et~al.} 1998, \aap, 340, L55

\bibitem[{{Sato} {et~al.}(1986){Sato}, {Nagase}, {Kawai}, {Kelley},
  {Rappaport}, \& {White}}]{Sato1986}
{Sato}, N., {Nagase}, F., {Kawai}, N., {et~al.} 1986, \apj, 304, 241

\bibitem[{{Sch{\"o}nherr} {et~al.}(2007){Sch{\"o}nherr}, {Wilms}, {Kretschmar},
  {Kreykenbohm}, {Santangelo}, {Rothschild}, {Coburn}, \&
  {Staubert}}]{Schoenherr2007}
{Sch{\"o}nherr}, G., {Wilms}, J., {Kretschmar}, P., {et~al.} 2007, \aap, 472,
  353

\bibitem[{{Schwarm}(2017)}]{Schwarm2017Thesis}
{Schwarm}, F.-W. 2017, PhD thesis, {Friedrich-Alexander-Universit\"at
  Erlangen-N\"urnberg}

\bibitem[{{Schwarm} {et~al.}(2017{\natexlab{a}}){Schwarm}, {Ballhausen},
  {Falkner}, {Sch{\"o}nherr}, {Pottschmidt}, {Wolff}, {Becker}, {F{\"u}rst},
  {Marcu-Cheatham}, {Hemphill}, {Sokolova-Lapa}, {Dauser}, {Klochkov},
  {Ferrigno}, \& {Wilms}}]{Schwarm2017b}
{Schwarm}, F.-W., {Ballhausen}, R., {Falkner}, S., {et~al.} 2017{\natexlab{a}},
  \aap, 601, A99

\bibitem[{{Schwarm} {et~al.}(2017{\natexlab{b}}){Schwarm}, {Sch{\"o}nherr},
  {Falkner}, {Pottschmidt}, {Wolff}, {Becker}, {Sokolova-Lapa}, {Klochkov},
  {Ferrigno}, {F{\"u}rst}, {Hemphill}, {Marcu-Cheatham}, {Dauser}, \&
  {Wilms}}]{Schwarm2017a}
{Schwarm}, F.-W., {Sch{\"o}nherr}, G., {Falkner}, S., {et~al.}
  2017{\natexlab{b}}, \aap, 597, A3

\bibitem[{{Sokolova-Lapa} {et~al.}(2021){Sokolova-Lapa}, {Gornostaev}, {Wilms},
  {Ballhausen}, {Falkner}, {Postnov}, {Thalhammer}, {F{\"u}rst}, {Garc{\'\i}a},
  {Shakura}, {Becker}, {Wolff}, {Pottschmidt}, {H{\"a}rer}, \&
  {Malacaria}}]{SokolovaLapa2021}
{Sokolova-Lapa}, E., {Gornostaev}, M., {Wilms}, J., {et~al.} 2021, \aap, 651,
  A12

\bibitem[{{Staubert} {et~al.}(2016){Staubert}, {Klochkov}, {Vybornov}, {Wilms},
  \& {Harrison}}]{Staubert2016}
{Staubert}, R., {Klochkov}, D., {Vybornov}, V., {Wilms}, J., \& {Harrison},
  F.~A. 2016, \aap, 590, A91

\bibitem[{{Staubert} {et~al.}(2014){Staubert}, {Klochkov}, {Wilms}, {Postnov},
  {Shakura}, {Rothschild}, {F{\"u}rst}, \& {Harrison}}]{Staubert2014}
{Staubert}, R., {Klochkov}, D., {Wilms}, J., {et~al.} 2014, \aap, 572, A119

\bibitem[{{Staubert} {et~al.}(2019){Staubert}, {Tr{\"u}mper}, {Kendziorra},
  {Klochkov}, {Postnov}, {Kretschmar}, {Pottschmidt}, {Haberl}, {Rothschild},
  {Santangelo}, {Wilms}, {Kreykenbohm}, \& {F{\"u}rst}}]{Staubert2019}
{Staubert}, R., {Tr{\"u}mper}, J., {Kendziorra}, E., {et~al.} 2019, \aap, 622,
  A61

\bibitem[{{Suchy} {et~al.}(2012){Suchy}, {F{\"u}rst}, {Pottschmidt},
  {Caballero}, {Kreykenbohm}, {Wilms}, {Markowitz}, \&
  {Rothschild}}]{Suchy2012}
{Suchy}, S., {F{\"u}rst}, F., {Pottschmidt}, K., {et~al.} 2012, \apj, 745, 124

\bibitem[{{Suleimanov} {et~al.}(2023){Suleimanov}, {Forsblom}, {Tsygankov},
  {Poutanen}, {Doroshenko}, {Doroshenko}, {Capitanio}, {Di Marco},
  {Gonz{\'a}lez-Caniulef}, {Heyl}, {La Monaca}, {Lutovinov}, {Molkov},
  {Malacaria}, {Mushtukov}, {Shtykovsky}, {Agudo}, {Antonelli}, {Bachetti},
  {Baldini}, {Baumgartner}, {Bellazzini}, {Bianchi}, {Bongiorno}, {Bonino},
  {Brez}, {Bucciantini}, {Castellano}, {Cavazzuti}, {Chen}, {Ciprini}, {Costa},
  {De Rosa}, {Del Monte}, {Di Gesu}, {Di Lalla}, {Donnarumma}, {Dov{\v{c}}iak},
  {Ehlert}, {Enoto}, {Evangelista}, {Fabiani}, {Ferrazzoli}, {Garcia}, {Gunji},
  {Hayashida}, {Iwakiri}, {Jorstad}, {Kaaret}, {Karas}, {Kislat}, {Kitaguchi},
  {Kolodziejczak}, {Krawczynski}, {Latronico}, {Liodakis}, {Maldera},
  {Manfreda}, {Marin}, {Marinucci}, {Marscher}, {Marshall}, {Massaro}, {Matt},
  {Mitsuishi}, {Mizuno}, {Muleri}, {Negro}, {Ng}, {O'Dell}, {Omodei},
  {Oppedisano}, {Papitto}, {Pavlov}, {Peirson}, {Perri}, {Pesce-Rollins},
  {Petrucci}, {Pilia}, {Possenti}, {Puccetti}, {Ramsey}, {Rankin}, {Ratheesh},
  {Roberts}, {Romani}, {Sgr{\`o}}, {Slane}, {Soffitta}, {Spandre}, {Swartz},
  {Tamagawa}, {Tavecchio}, {Taverna}, {Tawara}, {Tennant}, {Thomas}, {Tombesi},
  {Trois}, {Turolla}, {Vink}, {Weisskopf}, {Wu}, {Xie}, \&
  {Zane}}]{Suleimanov2023}
{Suleimanov}, V.~F., {Forsblom}, S.~V., {Tsygankov}, S.~S., {et~al.} 2023,
  \aap, 678, A119

\bibitem[{Tanaka(1986)}]{Tanaka1986}
Tanaka, Y. 1986, in Lecture Notes in Physics, Vol. 255, Radiation Hydrodynamics
  in Stars and Compact Objects, ed. D.~{Mihalas} \& K.-H.~A. {Winkler}
  (Springer Berlin, Heidelberg), 198

\bibitem[{{Tsygankov} {et~al.}(2019{\natexlab{a}}){Tsygankov}, {Doroshenko},
  {Mushtukov}, {Suleimanov}, {Lutovinov}, \& {Poutanen}}]{Tsygankov2019b}
{Tsygankov}, S.~S., {Doroshenko}, V., {Mushtukov}, A.~A., {et~al.}
  2019{\natexlab{a}}, \mnras, 487, L30

\bibitem[{{Tsygankov} {et~al.}(2006){Tsygankov}, {Lutovinov}, {Churazov}, \&
  {Sunyaev}}]{Tsygankov2006}
{Tsygankov}, S.~S., {Lutovinov}, A.~A., {Churazov}, E.~M., \& {Sunyaev}, R.~A.
  2006, \mnras, 371, 19

\bibitem[{{Tsygankov} {et~al.}(2007){Tsygankov}, {Lutovinov}, {Churazov}, \&
  {Sunyaev}}]{Tsygankov2007}
{Tsygankov}, S.~S., {Lutovinov}, A.~A., {Churazov}, E.~M., \& {Sunyaev}, R.~A.
  2007, Astron. Lett., 33, 368

\bibitem[{{Tsygankov} {et~al.}(2019{\natexlab{b}}){Tsygankov}, {Rouco
  Escorial}, {Suleimanov}, {Mushtukov}, {Doroshenko}, {Lutovinov}, {Wijnands},
  \& {Poutanen}}]{Tsygankov2019a}
{Tsygankov}, S.~S., {Rouco Escorial}, A., {Suleimanov}, V.~F., {et~al.}
  2019{\natexlab{b}}, \mnras, 483, L144

\bibitem[{{Verner} \& {Yakovlev}(1995)}]{Verner1995}
{Verner}, D.~A. \& {Yakovlev}, D.~G. 1995, \aaps, 109, 125

\bibitem[{{Vybornov} {et~al.}(2017){Vybornov}, {Klochkov}, {Gornostaev},
  {Postnov}, {Sokolova-Lapa}, {Staubert}, {Pottschmidt}, \&
  {Santangelo}}]{Vybornov2017}
{Vybornov}, V., {Klochkov}, D., {Gornostaev}, M., {et~al.} 2017, \aap, 601,
  A126

\bibitem[{{Wang}(2014)}]{Wang2014}
{Wang}, W. 2014, \mnras, 440, 1114

\bibitem[{Watanabe {et~al.}(2003)Watanabe, Sako, Ishida, Ishisaki, Kahn,
  Kohmura, Morita, Nagase, Paerels, \& Takahashi}]{Watanabe2003}
Watanabe, S., Sako, M., Ishida, M., {et~al.} 2003, ApJ, 597, L37

\bibitem[{{Watson} {et~al.}(1982){Watson}, {Warwick}, \& {Corbet}}]{Watson1982}
{Watson}, M.~G., {Warwick}, R.~S., \& {Corbet}, R.~H.~D. 1982, \mnras, 199, 915

\bibitem[{{White} {et~al.}(1976){White}, {Mason}, {Huckle}, {Charles}, \&
  {Sanford}}]{White1976}
{White}, N.~E., {Mason}, K.~O., {Huckle}, H.~E., {Charles}, P.~A., \&
  {Sanford}, P.~W. 1976, \apjl, 209, L119

\bibitem[{Wilms {et~al.}(2000)Wilms, Allen, \& McCray}]{Wilms2000}
Wilms, J., Allen, A., \& McCray, R. 2000, \apj, 542, 914

\bibitem[{{Yang} {et~al.}(2023){Yang}, {Wang}, {Liu}, {Chen}, {Wu}, {Tian}, \&
  {Chen}}]{Yang2023}
{Yang}, W., {Wang}, W., {Liu}, Q., {et~al.} 2023, \mnras, 519, 5402

\bibitem[{{Zhang} {et~al.}(2020){Zhang}, {Li}, {Lu}, {Song}, {Xu}, {Liu},
  {Chen}, {Cao}, {Bu}, {Chang}, {Chen}, {Chen}, {Chen}, {Chen}, {Chen}, {Cui},
  {Cui}, {Deng}, {Dong}, {Du}, {Fu}, {Gao}, {Gao}, {Gao}, {Ge}, {Gu}, {Guan},
  {Gungor}, {Guo}, {Han}, {Hu}, {Huang}, {Huo}, {Jia}, {Jiang}, {Jiang}, {Jin},
  {Jin}, {Li}, {Li}, {Li}, {Li}, {Li}, {Li}, {Li}, {Li}, {Li}, {Li}, {Li},
  {Liang}, {Liao}, {Liu}, {Liu}, {Liu}, {Liu}, {Liu}, {Liu}, {Lu}, {Lu}, {Luo},
  {Ma}, {Meng}, {Nang}, {Nie}, {Ou}, {Qu}, {Sai}, {Shang}, {Shen}, {Sun},
  {Tan}, {Tao}, {Tuo}, {Wang}, {Wang}, {Wang}, {Wang}, {Wang}, {Wang}, {Wang},
  {Wen}, {Wu}, {Wu}, {Wu}, {Xiao}, {Xiong}, {Yan}, {Yang}, {Yang}, {Yang},
  {Yi}, {Yuan}, {Zhang}, {Zhang}, {Zhang}, {Zhang}, {Zhang}, {Zhang}, {Zhang},
  {Zhang}, {Zhang}, {Zhang}, {Zhang}, {Zhang}, {Zhang}, {Zhang}, {Zhang},
  {Zhang}, {Zhang}, {Zhang}, {Zhang}, {Zhang}, {Zhao}, {Zhao}, {Zheng}, {Zhou},
  {Zhu}, {Zhu}, {Zhuang}, \& {Insight-HXMT Team}}]{Zhang2020}
{Zhang}, S.-N., {Li}, T., {Lu}, F., {et~al.} 2020, Sci. China Phys. Mech.
  Astron., 63, 249502

\end{thebibliography}
\appendix
\section{\textit{RXTE}/ASM-\bat cross-calibration}
\label{sec:AppendixA}
\begin{figure}
    \centering
    \resizebox{\hsize}{!}{\includegraphics{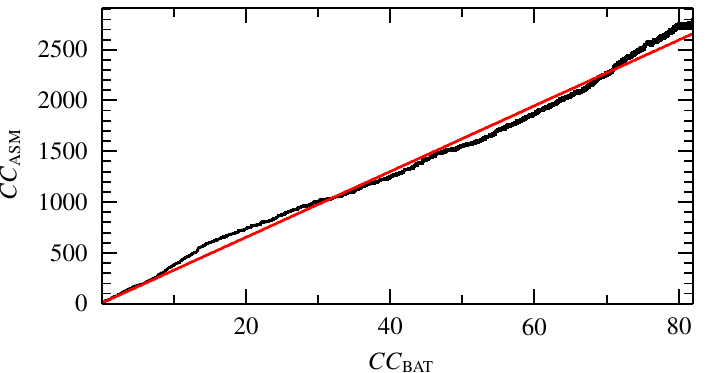}}
    \caption{Cross-calibration between \bat and \textit{RXTE}/ASM.
    Black: Cumulative sum of the ASM count rates as a function of the cumulative sum of the BAT count rates. Red: Fitted linear curve.}
    \label{fig:cumsum}
\end{figure}

In order to derive daily light curves for \gx{} for the past ${\sim}3$ decades, we use both \bat \citep{Krimm2013} and \textit{RXTE}/ASM \citep{levine1996FirstResultsAllSkyMonitorITAL} light curves binned to 1\,d time resolution\footnote{For \bat: \url{https://swift.gsfc.nasa.gov/results/transients/}. For \textit{RXTE}/ASM: \url{http://xte.mit.edu/asmlc/ASM.html}}. In order to combine the measurements from both observatories we aim to derive a simple method to relate their count rates. This is possible as there is a $\sim$6\,year time window in which both satellites observed the source almost daily. In total, we find 1900 days with coverage by both observatories between MJD 53416--55925. As the individual data points of the daily light curves can be strongly influenced by background radiation and noise they are not suited to derive a relation between the two detectors. We therefore calculate the cumulative sum of all data points for both light curves, i.e.,
\begin{equation}
    \mathrm{CC}_\mathrm{ASM}(T)=\sum_{i=1}^T C_\mathrm{ASM}(i),
\end{equation}
where $C_\mathrm{ASM}(i)$ is the count rate observed  with the All-Sky-Monitor on day $\mathrm{MJD}=i+53415$ and $T\in [1,2509]$. We show the obtained cumulative sum of the ASM as a function of the cumulative sum of the BAT monitor in Fig.~\ref{fig:cumsum}, together with a linear fit of the form
\begin{equation}
    \mathrm{CC}_\mathrm{ASM}(\mathrm{CC}_\mathrm{BAT})=m \cdot \mathrm{CC}_\mathrm{BAT} + t.
    \label{eq:cumsum}
\end{equation}
The best-fit parameters are given in  Table \ref{tab:cumsum}. The $\chi^2$ fit statistic does not take the uncertainties of the input values into account, which are used to calculate the expected values. However, as the uncertainties of $\mathrm{CC}_\mathrm{BAT}$ are very small in comparison to the corresponding quantity of the ASM, it is sufficient to not consider them for the fitting algorithm. We find that the fit describes the data very well, and that there are only very small deviations from the best-fit. 

\begin{table}
\caption{Best-fit parameters to fit function as in Eq. \eqref{eq:cumsum} and data in Fig. \ref{fig:cumsum}.}\label{tab:cumsum}
    \renewcommand{\arraystretch}{1.1}
    \begin{tabular}{lrr}
         \hline Parameter & Value & Unit  \\ \hline
         $m$& $32.395\pm0.021$ &  $CC_\mathrm{ASM}/CC_\mathrm{BAT}$\\
         $t$ & $4.98\pm0.25$ & $CC_\mathrm{ASM}$ \\ \hline
    \end{tabular}
\end{table}

Using Eq.~\eqref{eq:cumsum}, we can determine the BAT count rate from measured ASM count rates via
\begin{equation}
    C_\mathrm{BAT}=m^{-1} \cdot C_\mathrm{ASM}.
\end{equation}
The parameter $t$ is not required for this conversion as it corresponds to the - theoretically zero - offset between the two cumulative sums and bears no meaning for individual count rates. We can now use the above equation to estimate BAT count rates from ASM count rates for those MJDs where no BAT observation took place.

\begin{figure}
    \centering
    \resizebox{\hsize}{!}
    {\includegraphics{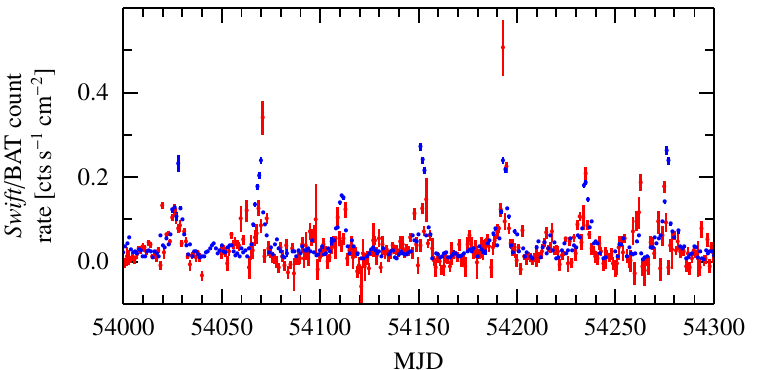}}
    \caption{Light curve comparison between BAT and converted ASM. Blue: Light curve of \gx{} as observed with \bat. Red: ASM light curve converted to BAT-comparable count rates.}
\label{fig:asm_plot}
\end{figure}

In order to judge how well this cross-calibration works, we show both BAT and ASM-converted BAT count rates for the interval MJD 54000--54300 in Fig.~\ref{fig:asm_plot}.
We note that the general orbital luminosity-dependency of \gx{} is apparent in the ASM-converted BAT rate, whereas there are a non-negligible number of outliers where the converted rate is significantly higher or lower than the BAT count rate. Overall, is evident that the procedure can be used to roughly estimate the corresponding \bat count rate from an observed ASM count rate; however it is also apparent that there are some data points in Fig.~\ref{fig:asm_plot} for which the values determined by the two observatories differ significantly. This is not surprising since instrument effects are not taken into account in the procedure and since ASM and \bat cover different energy bands of 2--10\,kev and 15--50\,keV, respectively. In general, the light curve obtained with \bat shows far fewer outliers and in comparison shows a more periodic light curve that more closely resembles the expected periodic X-ray emission that stems from the orbital characteristics of the system \citep[see, e.g.,][their Fig.~1]{Islam2014}. This is consistent with the harder energy band being less affected by processes outside of the accretion column like flaring or absorption.  Therefore, the \bat light curve is in our case favourable over the ASM light curve. In conclusion, we obtained a light curve combined from the two observatories, which roughly resembles the X-ray luminosity of the source.

\section{Gaussian soft emission}\label{sec:AppendixB}
\begin{figure}
    \centering
    \resizebox{\hsize}{!}{\includegraphics{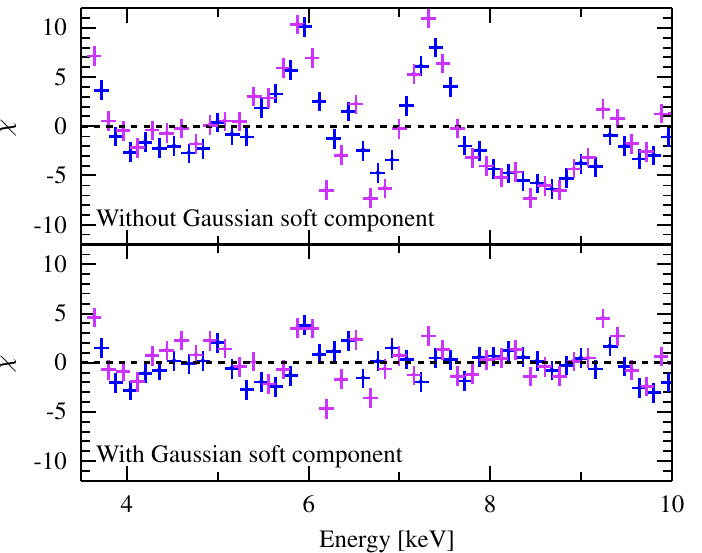}}
    \caption{$\chi$ residuals of \texttt{cutoffpl+bbody} fit with and without added Gaussian soft component. The component has a center position of ${\sim}$6.1\,keV, a width of ${\sim}$0.9\,keV and a normalization of ${\sim}0.07\, \mathrm{photons}\,\mathrm{s}^{-1}\,\mathrm{cm}^{-2}$. The color scheme for \nustar's focal plane modules is as in previous figures.}
    \label{fig:artificial}
\end{figure}

In Sect.~\ref{sec:wholeobs} we introduced a Gaussian soft emission component in our spectral model. In this appendix, we briefly explain why this parameter is required. Our preliminary spectral analysis includes the application of different phenomenological empirical models such as \texttt{FDcut}, \texttt{cutoffpl} and \texttt{NPEX}, the former two also in combination with blackbody components. Applying a comparably simple model of the form
\begin{align}
\begin{split}
f(E)=\texttt{detconst} \times \Bigl( &\texttt{PCF}(E) \times \texttt{continuum}(E)\\
&+\texttt{egauss}_\alpha(E)+\texttt{egauss}_\beta(E) \Bigr)
\end{split}
\end{align}
either with or without a gain shift shows very strong deviations from the data at energies of 6--7\,keV. The structure of the residuals indicate an incorrect modeling of the iron line complex, which can partly be solved by allowing the Fe\,K$\alpha$ line energy to shift to ${\sim}$6.1\,keV. Such a shift is not expected from a physical point of view; further it does not fully solve the modeling issues. In order obtain a better empirical description of the spectrum in the range 6--7\,keV, we introduce a Gaussian emission component centered at ${\sim}$6\,keV. We show the residuals of a \texttt{cutoffpl+bbody} fit with and without the Gaussian emission component in the soft X-ray band in Fig.~\ref{fig:artificial}.

We note that the additional soft emission component is very prominent and contributes around 20\% of the total model counts at its centroid energy. Whilst the X-ray spectrum of \gx is known to exhibit a Compton shoulder \citep[see, e.g.,][Fig.~1]{Watanabe2003} at energies of ${\sim}$6.2\,keV, the residuals shown in the upper panel of Fig.~\ref{fig:artificial} do not indicate that the Compton shoulder is responsible for the observed deviations from the model. We argue that both calibration issues and a complex continuum shape in the soft X-ray band lead to deviations from the model, which therefore needs to include the empirical Gaussian emission component. Furthermore, \nustar{}'s energy resolution of ${\sim}$400\,eV at 6.4\,keV \citep[see][]{Harrison2013} does not suffice to model the soft emission in detail. To model both the soft and the hard X-ray band simultaneously, coordinated contemporaneous observations of \nustar{} and \textit{XMM}-Newton or \textit{Swift}/XRT are required.

In the present work, we see the inclusion of the additional soft Gaussian feature as a means to account for the complex soft X-ray emission structure of \gx, which is not of interest in this work and does not influence our investigation of the cyclotron lines.

\section{Detection significance of the 35\,keV CRSF in the phase-resolved spectra}
Based on the fact that the strength of the 35\,keV CRSF in the phase-resolved spectral fits (see Fig.~\ref{fig:prs_phDep}) is predominantly low and under ${\sim}$5\,keV, the question arises with which significance the line is detected and whether even spectral fits without it can provide a good description of the data. We therefore employ the same statistical test as for CRSF2 as in Sect.~\ref{sec:phaseresolved} in order to determine the detection confidence for each phase bin, respectively. Analogous to Table~\ref{tab:prsCRSFConfidence}, we give the results of this statistical test in Table~\ref{tab:prsCRSF1Confidence}.

\begin{table*}
    \caption{Summary of search for CRSF1 in the phase-resolved spectra.}
    \begin{tabular}{cccrrrrc}
    \hline \multirow{2}{*}{Phase Bin} & \multirow{2}{*}{Residuals indicate CRSF1 \tablefootmark{a)}} & \multirow{2}{*}{$E,\sigma$ constrained\tablefootmark{b)}} & \multirow{2}{*}{$\Delta \chi^2$\,\tablefootmark{c)}} & \multirow{2}{*}{$N_\mathrm{FP}$ \tablefootmark{d)}} & \multicolumn{2}{c}{Confidence} & \multirow{2}{*}{Final evaluation\tablefootmark{e)}} \\
    &&&&& [\%] & [$\sigma$] & \\\hline
    0 & Y & Y & 21 & 15 & 99.985 & 3.7$\sigma$ & Y \\
1 & Y & Y & 90 & 0 & >99.999 & >4.4$\sigma$ & Y \\
2 & N & Y & 53 & 0 & >99.999 & >4.4$\sigma$ & Y \\
3 & Y & Y & 45 & 0 & >99.999 & >4.4$\sigma$ & Y \\
4 & Y & Y & 101 & 0 & >99.999 & >4.4$\sigma$ & Y \\
5 & N & Y & 20 & 36 & 99.964 & 3.5$\sigma$ & Y \\
6 & Y & Y & 92 & 0 & >99.999 & >4.4$\sigma$ & Y \\
7 & N & Y & 25 & 1 & 99.999 & 4.4$\sigma$ & Y \\ \hline     \end{tabular} \\
    \tablefoottext{a}{Indicator whether the residuals of a fit with excluding CRSF1 indicate a missing model component at 35\,keV.} \\
    \tablefoottext{b}{Can the energy, width and strength of CRSF1 be constrained in a spectral fit?}\\
    \tablefoottext{c}{Improvement in statistic upon inclusion of CRSF1 in the spectral model.}\\
    \tablefoottext{d}{Number of false positives out of 100\,000 simulated spectra where $\Delta \chi^2$ exceeds the corresponding value upon inclusion of CRSF1.}\\
    \tablefoottext{e}{Final evaluation whether CRSF1 is detected in the corresponding phase bin.}
    \label{tab:prsCRSF1Confidence}
\end{table*}

It is evident that the 35\,keV CRSF can be detected with at least 3$\sigma$ significance during the whole pulse phase. We consequently argue that CRSF1 is present throughout the whole pulse phase as found in many previous studies \citep[see, e.g.,][]{Kreykenbohm2004,Fuerst2018}

\end{document}